\documentclass[12pt]{article}

\usepackage{cmap}
\usepackage{indentfirst} 
\usepackage{epsfig}

\usepackage{amsmath}
\usepackage{amssymb}
\usepackage{mathrsfs}
\usepackage{mathabx}
\usepackage{slashed}
\usepackage{nicefrac}
\usepackage{array}
\usepackage{color}

\newcommand{\bra}[1]{\big\langle \, #1\,\big\vert}
\newcommand{\ket}[1]{\big\vert\, #1\,\big\rangle}
\newcommand{\bracket}[2]{\big\langle \, #1 \big\vert \, #2\,\big\rangle}

\newcommand{\matrel}[3]{\bra{#1} #2\ket{#3}}

\begin{document}

\begin{center}
{\bf {\Large Wigner Inequalities for Test of Hypothesis of Realism and Concepts of Macroscopic and Local Realism}}
\end{center}

\vspace{0.7cm}
\begin{center}
{\large N.~Nikitin, K.~Toms}
\end{center}


\subsection*{Abstract}
We propose a new Wigner inequality suitable for test of the hypothesis of realism. We show that this inequality is not identical neither to the
well-known Wigner inequality nor to the Leggett-Garg inequality in Wigner form. The obtained inequality is suitable for test of realism not only
in quantum mechanical systems, but also in quantum field systems.

Also we propose a mathematically consistent derivation of the Leggett--Garg inequality in Wigner form, which was recently presented in the literature,
for three and $n$ distinct moments of time. Contrary to these works, our rigor derivation uses Kolmogorov axiomatics of probability theory.
We pay special attention to the construction and studies of the spaces of elementary outcomes. 
Basing on the the Leggett--Garg inequality in Wigner form for $n$ distinct moments of time we prove that any unitary evolution of a quantum 
system contradicts the concept of macroscopic realism. We show that application of the concept of macroscopic realism to any quantum system
leads to ``freezing'' of the system in the initial state.

It is shown that for a particle with an infinite number of observables the probability to find a pair of the observables in some defined state is zero,
even if the operators of these observables commute. This fact might serve as an additional logical argument for the contradiction between quantum
theory and classical realism.


\section*{Introduction}

It seems that foundations of the concept of local realism (LR) were used by Einstein while creating Special Relativity theory. But consistently the concept of LR was introduced in the famous paper by Einstein-Podolsky-Rosen \cite{Einstein:1935rr}. Local realism comprises the following three statements.

1) Classic realism: an aggregate of all physical characteristics (in classical terms) of a system exists jointly and is independent of an observer, even if the observer cannot simultaneously measure these characteristics with any classical measurement device. 

2) Locality: if two measurements are performed in spatially-separated points of the spacetime, then the readings of one classical device do not affect the readings of a second one in any way.

3) Freedom of choice: the observer can freely choose any experimental parameters from the available ones.

In their pioneering paper \cite{Leggett:1985zz}  Leggett and Garg have rendered concrete the intuitive notions of properties of classical objects using two simple principles: the ``Macroscopic realism per se'' and ``Non-invasive measurability''. Using these principles the authors have suggested inequalities (LGI), which are satisfied for any physical system that follows our``macroscopic intuition''.
Typically two canonical principles are combined with a third one, the ``Induction'' \cite{Leggett:2002,Leggett:2008,Kofler:2008}. Jointly these three principles are called ``Macroscopic realism'' or ``Macrorealism''.

Let us formulate these three principles:

1) Macroscopic realism per se: a physical system which can obtain several macroscopically distinct states exists in one and only one of its possible states at any time.

2) Non-invasive measurability principle: it is possible to determine the state of a physical system while introducing only a negligible impact on its further dynamics.

3) Induction: reflects a layman's understanding of the freedom of will, i.e. that the result of the current measurement does not influence what measurements the observer will perform in the future.

Both macroscopic realism per se and non-invasive measurability are satisfied in classical physics and are violated in the quantum paradigm. First -- because of quantum superposition. Second -- because, according to the Bohr or Dirac--von Neumann projection postulates, a state vector or a density matrix of a quantum system is subject to reduction when measured with a classical device. The induction principle is satisfied in both the quantum and the classical worlds. This principle is closely linked with the freedom of will principle (third local realism condition) and with the No-signaling condition \cite{Cirelson:1980,PRboxes:1984}.  

Usually the no-signaling condition is written in the following form \cite{PRboxes:1984}:
\begin{eqnarray}
\label{NSC}
\sum\limits_a\, w(a,\, b_\beta,\, \ldots\, |\, A,\, B,\, \ldots)\, =\, w(b_\beta,\,\ldots\, |\, B,\, \ldots),
\end{eqnarray}
where $A$ is an observable selected for measurement, $a$ is the measured value of the observable $A$, and $\sum\limits_a$ sums all possible values of the observable $A$. Often $A$ is thought of as a state of a classical device which measures the corresponding observable. The same notation is used for the observable $B$. For two or more spatially-separated measurement devices in classical paradigm, the no-signaling condition is a corollary of the locality of special relativity. In non-relativistic quantum mechanics, the NCS is known as Eberhard theorem and is a corollary of probabilistic nature of the outcome of any measurement of a quantum system with a classical device. In this case it is supposed that the quantum system is separated into few subsystems. The observable $A$ is related to one of these subsystems together with its macro-device. The observable $B$ and its macro-device are related to some other subsystem. In the formalism of Popescu–Rohrlich boxes (PR--boxes) no-signaling condition is introduced as one of the axioms \cite{PRboxes:1984}.

Using classic realism and the no-signaling condition it is possbile to obtain the well-known Wigner inequalities \cite{wigner}. The details are outlined in Appendix~\ref{sec:Awlg}. They are important for the comparison below of Wigner inequalities and various forms of Leggett--Garg inequalities considered here. Note that a delicate question arises here: to what extent the NSC is equivalent to the condition of locality of the LR concept. Here we assume that NSC follows directly from the condition of locality.

The Non-invasive measurability principle in the concept of macroscopic realism for the derivation of LGI plays the same role as the principle of locality in the concept of local realism for the derivation of Bell inequalities \cite{Bell:1964kc,Bell:1964fg,Clauser:1969ny}.   

The ``No-signaling in time'' condition was introduced in \cite{PhysRevA.87.052115}. This condition may be considered as an analog of the no-signaling condition for LGI and as an alternative statistical version of non-invasive measurability. No-signaling in time demands that the probability $w(q_j,\, q_i,\, \ldots\, |\, t_j,\, t_i,\,\ldots)$ of measurement of an observable $Q$ at times $t_i$, $t_j > t_i$ and so on, does not depend on the state of the observable $Q$ at time $t_k \ne\{ t_i,\, t_j,\, \ldots \}$. Denoting $Q(t_i)$ as $q_i$, no-signaling in time condition may be written as follows.
\begin{eqnarray}
\label{NSIT}
\sum\limits_{q_k}\, w(q_j,\, q_k,\, q_i,\, \ldots\, |\, t_j,\, t_k,\, t_i,\, \ldots )\, =\, w(q_j,\, q_i,\, \ldots\, |t_j,\, t_k,\, t_i,\, \ldots)\,\equiv\, w(q_j,\, q_i,\, \ldots\, |t_j,\, t_i,\, \ldots).
\end{eqnarray}
In this form the analogy between (\ref{NSC}) and (\ref{NSIT}) is quite obvious. Note that the no-signaling condition and no-signaling in time condition are satisfied in the classical paradigm. However the no-signaling condition is naturally obtainable from quantum mechanics \cite{eberhard}, в while no-signaling in time is not \cite{PhysRevA.87.052115}. Note, that the role of NSIT in obtaining the relations testing the MR concept may be more complex than the role of NSC for the LR concept. For example in \cite{PhysRevA.96.012121} it was shown that there are various necessary and sufficient conditions for the MR concept depending on the chosen NIM form.

A test of the Leggett--Garg inequalities requires the technique of non-invasive (soft) measurements. However, if we go from a non-invasive measurement of an observable $Q(t)$ of a single particle at distinct times to a fully-invasive measurement of fully correlated observables of a pair of particles ``$1$'' and ``$2$'', for instance $Q^{(1)}(t)$ and $Q^{(2)}(t)$, at two distinct times, it is possible to obtain an inequality similar to the LGI for one particle, but without using non-invasive measurability. What statement could be tested in violation of such an inequality? There is no common opinion in the literature \cite{PhysRevA.54.1798, Gangopadhyay:2013aha, Formaggio:2016cuh, Naikoo:2018vug}. Most often this statement is the hypothesis of realism \cite{Formaggio:2016cuh, Pusey:2011de}. This hypothesis is a peculiar mix of classic realism and macroscopic realism per se, but not their mechanical union. 

In the current work we will use the following formulation of hypothesis of realism:

1) At any time $t_i$ a system is in a``real physical state'' which exists impartially and independently of any observer.``Real physical states'' are distinguished from each other by the values of observables that characterize the system under study. We do not suppose these values to be jointly measurable by any macroscopic device.

2) Observable physical states of a system are distinguished by the values of variables which can be jointly measurable in the system at time $t_i$. We will equate``real physical state'' and ``ontic state'' \cite{Pusey:2011de,Harrigan2010}, describing it using joint probabilities of the observable states.

3) For the considered system the no-signaling in time condition in form (\ref{NSIT}) and/or no signalling conditions are hold.

4) The experimentalist has free will to plan, perform, and analyze the results of the experiments on the system.

Note that hypothesis of realism with addition of NSC/NSIT and the free will condition may serve as a suitable base for obtaining relations which are true in classical physics, but are not true in the framework of quantum field theory (QFT). Actually, in derivation of Bell or Wigner inequalities a static correlated state is assumed. Such approximation is valid for non-relativistic quantum mechanics, but in the framework of QFT it is in principle not possible to exclude interaction of the fields related to the correlated state with other particle fields and with vacuum fluctuations. Such interactions may decrease correlation over time \cite{KAZAKOV20122914}. However the time dependence is not included into the concept of local realism. In the framework of macroscopic realism the time dependence is introduced, however it is severely restricted by the NIM condition. Also in MR the locality, which is one of the main properties of QFT, is not considered.

This paper is a continuation of series of papers studying time-dependent extensions of the Wigner inequalities \cite{Nikitin:2009sr,Nikitin:2014pqa,Nikitin:2015bca,Nikitin:2016ilm}. This series was stimulated by a desire to generalize the Bell inequalities \cite{Bell:1964kc,Bell:1964fg,Clauser:1969ny} and Wigner inequalities \cite{wigner} for quantum field theory (QFT). Given that in the formalism of QFT the probability calculation procedure is well defined (contrary to the correlator calculation procedure), the Wigner inequalities are preferable for their generalization in QFT. However in QFT it is not possible to use the well-known static (i.e. time independent) form of the Wigner inequalities \cite{wigner}, because field interactions and interactions with vacuum fluctuations cannot be neglected. In references \cite{Nikitin:2009sr,Nikitin:2014pqa,Nikitin:2015bca} some attempts have been made to obtain a non-stationary version of the Wigner inequalities in the framework of local realism. In \cite{Nikitin:2016ilm} another attempt has been made: using the Bayes theorem and its combination with hypothesis of realism. The results of these studies have demonstrated that the description of testing of quantum concepts in the ``probabilistic approach'' (which is based on ideas of Wigner) has potential comparable to the commonly used ``orthodox'' correlator-based approach. 
This fact has stimulated the authors to make in the current paper an attempt to apply the Wigner formalism to Leggett--Garg inequalities.

The Leggett--Garg inequalities in Wigner form for three and $n$ moments of time were first introduced in \cite{Saha:2015}. The macroscopic realism concept was used in the derivation. However in \cite{Saha:2015} and in consequent works \cite{Mal:2016,Das:2018} there has no study been performed of the spaces of the elementaty outcomes and structure of events, that correspond to the probabilities from the Leggett--Garg inequalities in Wigner form. We will show that such a study may lead to some non-trivial statements about the scope of the application of the Leggett--Garg inequalities in Wigner form and for analogous inequalities suitable for test of hypothesis of realism.

Another interesting development was introduced in \cite{Kumari:2017}, where the role of the NIST condition has been studied and various situations for violation of the LGI have been considered.

The paper is organized as follows. In the Introduction section we present definitions of local realism, macroscopic realism, and hypothesis of realism, necessary for our derivations. In Section~\ref{sec:LGI-01} basing on Kolmogorov axiomatics some variants of Leggett-Garg inequalities in Wigner form are derived for three distinct times. We discuss there differences in spaces of elementary outcomes for the inequalities considered. Also we demonstrate violation of the obtained inequalities in quantum theory. 
Section~\ref{sec:LGI-02} is devoted to generalization of the Leggett--Garg inequalities in Wigner form for $n$ moments of time. A theorem is proven that if a quantum mechanical transition is compatible with the macroscopic realism concept then the probability of any quantum transition is zero. In Section~\ref{sec:LGI-02a} we consider new foundation of incompatibility of local realism with quantum theory.  In Section~\ref{sec:LGI-03} we propose an inequality for testing the hypothesis of realism and a difference between it and the Wigner inequality is shown. The Conclusion section contains the main results of the present work. Appendix~\ref{sec:Awlg} contains a short discussion of Wigner inequalities as background information. In Appendix~\ref{sec:B-time-evolution} a technique for calculation of time evolution of neutral pseudoscalar $B$-mesons is presened.


\section{Derivation of Leggett--Garg inequalities in Wigner form for three distinct times}
\label{sec:LGI-01}

We use the macroscopic realism concept and Kolmogorov axiomatics of probability theory for the derivation. According to macroscopic realism per se, a dichotomic observable $Q$ at any time $t_i$ may have one and only one of its possible values $Q(t_i) \equiv q_i = \pm 1$. Let us consider triple probabilities $w(q_k,\, q_j,\, q_i\, |t_j,\, t_i)$,  $w(q_j,\, q_k,\, q_i\, |t_j,\, t_i)$, $w(q_j,\, q_i,\, q_k\, |t_j,\, t_i)$, and so on, where $k \ne \{i,\, j\}$. We suppose that at times $t_i$ and $t_j \ne t_i$ the measurement of the observable $Q$ has taken place, while there has been no measurement of $Q$ at time $t_k$, yet at that time $Q$ had some defined value (according to macroscopic realism per se). Taking into account the non-invasive measurability condition one can see that it is not important at which two of three times the measurements of $Q(t)$ have been made. Hence we can write $w(q_k,\, q_j,\, q_i\, |t_k,\, t_j,\, t_i)$ and $w(q_j,\, q_k,\, q_i\, |t_j,\, t_k,\, t_i)$ instead of $w(q_k,\, q_j,\, q_i\, |t_j,\, t_i)$ and $w(q_j,\, q_k,\, q_i\, |t_j,\, t_i)$.

Let us consider three distinct times $t_3 > t_2 > t_1$, and the observable $Q$ is measured at any two of them. Then in the triple probabilities the values $q_1$, $q_2$, and $q_3$ present only once. The order is not important. We will present the values of the observable $Q$ from right to left ordered by time. Let us apply the same rule to double probabilities. That is, we will deal with triple probabilities like $w(q_3,\, q_2,\, q_1\, |t_3,\, t_2,\, t_1)$ or with double ones like $w(q_3,\, q_1\, |\, t_3,\, t_2,\, t_1)$ and so on. Also if $q_i = \pm 1$ we denote it as $q_{i\, \pm}$. The latter is introduced in order to clarify the link between the derivation of the Leggett--Garg inequality in Wigner form and the derivation of the well-known Wigner inequality for a single particle (see Appendix~\ref{sec:Awlg}).

Let us denote the space $\Omega^{(LG)}$ of the elementary outcomes $\omega_{i j}^{(LG)} \in \Omega^{(LG)}$ as consisting of the aggregate of $\{ q_{3\,\alpha},\, q_{2\,\beta},\, q_{1\,\gamma}\}$, where $\{\alpha,\, \beta,\,\gamma \} = \{+,\, -\}$. Note, that under the non-invasive measurability conditions, times $t_i$ and $t_j$, when the measurement has taken place for the observable $Q(t)$, do not enter the definition of $\Omega^{(LG)}$. The structure of the $\Omega^{(LG)}$ space repeats the structure of the $\Omega$ space, which has been used for the derivation of the Wigner inequality (\ref{wigner_takoy_wigner}) for a single particle (see Appendix~\ref{sec:Awlg}).
In the $\Omega^{(LG)}$ space let us denote an elementary event ${\cal K}^{(LG)}_{q_{3\,\alpha},\, q_{2\,\beta},\, q_{1\,\gamma}} \subseteq \Omega$, that at times $t_3$, $t_2$, and $t_1$ the observable $Q$ has been in states $q_{3\,\alpha}$, $q_{2\,\beta}$, and $q_{1\,\gamma}$ accordingly. Non-invasive measurability conditions tell us that it is not important in which two of the three moments of time the measurement took place. The set of elementary events ${\cal K}^{(LG)}_{q_{3\,\alpha},\, q_{2\,\beta},\, q_{1\,\gamma}}$ forms $\sigma$--algebra ${\cal F}^{(LG)}$, with structure isomorph to $\sigma$-algebra ${\cal F}$ from Appendix~\ref{sec:Awlg}. On $\left ( \Omega^{(LG)},\, {\cal F}^{(LG)}\right )$ let us introduce a real non-negative $\sigma$--additive measure $w(\ldots\, |\,\, t_3,\, t_2,\, t_1 )$. Triplet $\Omega^{(LG)}$, ${\cal F}^{(LG)},\, w(\ldots\, |\,\, t_3,\, t_2,\, t_1 )$ forms a probabilistic model of the task. It is obvious that this is a model related to classical physics.

We introduce the events 
\begin{eqnarray}
\label{eventsKforLG1part}
&&{\cal K}^{(LG)}_{32} = {\cal K}^{(LG)}_{\, q_{3\, +},\, q_{2\, -},\, q_{1\, +}}\cup\,{\cal K}^{(LG)}_{\, q_{3\, +},\, q_{2\, -},\, q_{1\, -}},\quad
{\cal K}^{(LG)}_{21} = {\cal K}^{(LG)}_{\, q_{3\, +},\, q_{2\, -},\, q_{1\, +}}\cup\,{\cal K}^{(LG)}_{\, q_{3\, -},\, q_{2\, -},\, q_{1\, +}}, \nonumber\\
&&{\cal K}^{(LG)}_{31} = {\cal K}^{(LG)}_{\, q_{3\, +},\, q_{2\, +},\, q_{1\, -}}\cup\,{\cal K}^{(LG)}_{\, q_{3\, +},\, q_{2\, -},\, q_{1\, -}}.
\end{eqnarray}
and elementary outcomes $\omega^{(LG)}_{32}$,  $\omega^{(LG)}_{21}$, and $\omega^{(LG)}_{31}$, which correspond to events ${\cal K}^{(LG)}_{32}$, ${\cal K}^{(LG)}_{21}$, and ${\cal K}^{(LG)}_{31}$ accordingly. Different values of the observable $Q$ at any time are independent events, hence 
\begin{eqnarray}
\label{w_q1q2q3}
w(q_{3\, +},\, q_{2\, -}\, |\, t_3,\, t_2,\, t_1) &=&
\sum\limits_{\omega^{(LG)}_{32} \in\, {\cal K}^{(LG)}_{32}}\,\sum\limits_{q_1}\, w(\omega^{(LG)}_{32},\, q_{3\, +},\, q_{2\, -},\, q_1\, |\, t_3,\, t_2,\, t_1); \nonumber  \\
w(q_{2\, -},\, q_{1\, +}\, |\, t_3,\, t_2,\, t_1) &=&
\sum\limits_{\omega^{(LG)}_{21} \in\, {\cal K}^{(LG)}_{21}}\,\sum\limits_{q_3}\, w(\omega^{(LG)}_{21},\, q_3,\, q_{2\, -},\, q_{1\, +}\, |\, t_3,\, t_2,\, t_1); \\ 
w(q_{3\, +},\, q_{1\, -}\, |\, t_3,\, t_2,\, t_1) &=&
\sum\limits_{\omega^{(LG)}_{31} \in\, {\cal K}^{(LG)}_{31}}\,\sum\limits_{q_2}\, w(\omega^{(LG)}_{31},\, q_{3\, +},\, q_2,\, q_{1\, +}\, |\, t_3,\, t_2,\, t_1),  \nonumber  
\end{eqnarray}
where $\sum\limits_{q_i}$ means $\sum\limits_{q_i = -1}^{+1}$. The sum of the probabilities $w(q_{2\, -},\, q_{1\, +}\, |\, t_3,\, t_2,\, t_1)$ and $w(q_{3\, +},\, q_{1\, -}\, |\, t_3,\, t_2,\, t_1)$ is defined on a set ${\cal K}^{(LG)}_{321} = {\cal K}^{(LG)}_{21} \cup {\cal K}^{(LG)}_{31}$. And ${\cal K}^{(LG)}_{32}\subseteq {\cal K}^{(LG)}_{321}$; this follows from (\ref{eventsKforLG1part}) . Hence all the elementary outcomes belong to the set ${\cal K}^{(LG)}_{321}$, i.e. $\{\omega^{(LG)}_{21},\, \omega^{(LG)}_{31},\,\omega^{(LG)}_{32}\}\,\in\, {\cal K}^{(LG)}_{321}$. Taking into account that all the probabilities that enter the sums (\ref{w_q1q2q3}) are non-negative,  we obtain the Leggett--Garg inequality in Wigner form for a single particle:
\begin{eqnarray}
\label{LGW-1part}
w(q_{3\, +},\, q_{2\, -}\, |\, t_3,\, t_2,\, t_1)\,\le\,
w(q_{2\, -},\, q_{1\, +}\, |\, t_3,\, t_2,\, t_1)\, +\, w(q_{3\, +},\, q_{1\, -}\, |\, t_3,\, t_2,\, t_1),
\end{eqnarray}
which is defined on the set ${\cal K}^{(LG)}_{321} \subseteq \Omega^{(LG)}$. 
The Leggett--Garg inequality in Wigner form (\ref{LGW-1part}), which is obtained in the framework of macroscopic realism and classical probability theory, is analogous to Wigner inequality (\ref{wigner_takoy_wigner}) for a single particle, obtained under classic realism condition and Kolmogorov axiomatics of probability theory. We stress that the no-signaling in time condition (\ref{NSIT}) was not used for the derivation of  (\ref{LGW-1part}). The sums (\ref{w_q1q2q3}) are one of the properties of independent events in the Kolmogorov formalism of probability theory and are not a corollary of the no-signaling in time condition. Here one can see another full analogy with the substantiation of formula (\ref{w_ABC}). The possibility to consider the events with different values of $q_i$ as independent follows from macroscopic realism per se and non-invasive measurability. Let us emphasize the fact that neither macroscopic realism per se nor non-invasive measurability do not allow to perform summation  (\ref{w_q1q2q3}). In order to substantiate the summation it is necessary to build the probabilistic model ($\Omega^{(LG)}$, ${\cal F}^{(LG)},\, w(\ldots\, |\,\, t_3,\, t_2,\, t_1 )$) of the considered task. Introduction of such a model and its use for the derivation of (\ref{LGW-1part}) is the first important result of the present work. It distincts the derivation of (\ref{LGW-1part}) from the analogous inequality in \cite{Saha:2015,Kumari:2017}.

For the three times $t_1$, $t_2$, and $t_3$ in the framework of macroscopic realism it is possible to write yet another type of the Leggett--Garg inequalities in Wigner form. Different from the inequality (\ref{LGW-1part}), this one is defined on the space of states $\Omega^{(LG)}$ and cannot be defined on its subsets. 
Let us introduce the events: 
\begin{eqnarray}
{\cal K}^{(LG)}_{1+}& =& 
                                          {\cal K}^{(LG)}_{\, q_{3\, +},\, q_{2\, +},\, q_{1\, +}}\cup\,{\cal K}^{(LG)}_{\, q_{3\, +},\, q_{2\, -},\, q_{1\,+}}\cup\,
                                          {\cal K}^{(LG)}_{\, q_{3\, -},\, q_{2\, +},\, q_{1\, +}}\cup\,{\cal K}^{(LG)}_{\, q_{3\, -},\, q_{2\, -},\, q_{1\, +}}, \nonumber \\
{\cal K}^{(LG)}_{1-} &=& 
                                          {\cal K}^{(LG)}_{\, q_{3\, +},\, q_{2\, +},\, q_{1\, -}}\cup\,{\cal K}^{(LG)}_{\, q_{3\, +},\, q_{2\, -},\, q_{1\,-}}\cup\,
                                          {\cal K}^{(LG)}_{\, q_{3\, -},\, q_{2\, +},\, q_{1\, -}}\cup\,{\cal K}^{(LG)}_{\, q_{3\, -},\, q_{2\, -},\, q_{1\, -}}
\nonumber
\end{eqnarray}
and define probabilities
\begin{eqnarray}
w(q_{1\, +},\, |\, t_3,\, t_2,\, t_1) &=&
\sum\limits_{\omega^{(LG)}_{1+} \in\, {\cal K}^{(LG)}_{1+}}\,\sum\limits_{q_3}\,,\sum\limits_{q_2}\, 
w(\omega^{(LG)}_{1+},\, q_3,\, q_2,\, q_{1\, +}\, |\, t_3,\, t_2,\, t_1); \nonumber  \\
w(q_{1\, -},\, |\, t_3,\, t_2,\, t_1) &=&
\sum\limits_{\omega^{(LG)}_{1-} \in\, {\cal K}^{(LG)}_{1-}}\,\sum\limits_{q_3}\,,\sum\limits_{q_2}\, 
w(\omega^{(LG)}_{1-},\, q_3,\, q_2,\, q_{1\, -}\, |\, t_3,\, t_2,\, t_1). \nonumber
\end{eqnarray}
Due to the normalization condition, the sum
$
w(q_{1\, +},\, |\, t_3,\, t_2,\, t_1) + w(q_{1\, -},\, |\, t_3,\, t_2,\, t_1) = 1
$
and is defined on the set ${\cal K}^{(LG)}_{1+} \cup {\cal K}^{(LG)}_{1-} \equiv \Omega^{(LG)}$. Taking into account the relations
\begin{eqnarray}
w(q_{2\, +},\, q_{1\, +}\, |\, t_3,\, t_2,\, t_1)\, +\, w(q_{2\, -},\, q_{1\, +}\, |\, t_3,\, t_2,\, t_1) &=& w(q_{1\, +}\, |\, t_3,\, t_2,\, t_1); \nonumber \\
w(q_{3\, +},\, q_{1\, -}\, |\, t_3,\, t_2,\, t_1)\, +\, w(q_{3\, -},\, q_{1\, -}\, |\, t_3,\, t_2,\, t_1) &=& w(q_{1\, -}\, |\, t_3,\, t_2,\, t_1) \nonumber 
\end{eqnarray}
from (\ref{LGW-1part}) we derive 
\begin{eqnarray}
\label{LGW-1part-NEW}
w(q_{3\, +},\, q_{2\, -}\, |\, t_3,\, t_2,\, t_1)\, +\, 
w(q_{2\, +},\, q_{1\, +}\, |\, t_3,\, t_2,\, t_1)\, +\,
w(q_{3\, -},\, q_{1\, -}\, |\, t_3,\, t_2,\, t_1)\, \le\, 1,
\end{eqnarray}
which is defined on the space of states $\Omega^{(LG)}$, contrary to (\ref{LGW-1part}). The inequality (\ref{LGW-1part-NEW}) has a direct analog among various forms of Wigner inequality. This analog is obtained in Appendix A of \cite{Nikitin:2009sr}.


\section{Generalization of the Leggett--Garg inequality in Wigner form for $n$ times}
\label{sec:LGI-02}

The inequality (\ref{LGW-1part}) can be generalized for $n$ times $t_n > t_{n -1} > \ldots > t_1$. Let us introduce a space of states $\Omega^{(LGn)}$, which consist of the aggregates $\{q_{n\,\alpha},\,q_{n-1\,\beta}\ldots,\, q_{2\,\gamma},\, q_{1\,\delta}\}$, where the Greek indices are $+$ or $-$.  On $\Omega^{(LGn)}$ we define elementary events ${\cal K}^{(LGn)}_{q_{n\,\alpha},\,q_{n-1\,\beta}\ldots,\, q_{2\,\gamma},\, q_{1\,\delta}} \subseteq \Omega$, $\sigma$--algebra ${\cal F}^{(LGn)}$, and $\sigma$--additive probability measure $w(\ldots\, |\, t_n,\, t_{n-1},\,\ldots\,\, t_1)$. Analogs of events ${\cal K}^{(LG)}_{ij}$ in this space are more complicated. For example, the analog of ${\cal K}^{(LG)}_{32}$ will have a form:
\begin{eqnarray}
{\cal K}^{(LGn)}_{32} &=& {\cal K}^{(LGn)}_{\,q_{n\, +},\, q_{(n-1)\, +}\,\ldots\,  q_{3\, +},\, q_{2\, -},\, q_{1\, +}}\cup\,
                                           {\cal K}^{(LGn)}_{\,q_{n\, -},\, q_{(n-1)\, +}\,\ldots\,  q_{3\, +},\, q_{2\, -},\, q_{1\, +}}\cup\,\ldots\nonumber \\
                   &\ldots&\cup\,{\cal K}^{(LGn)}_{\,q_{n\, -},\, q_{(n-1)\, -}\,\ldots\,  q_{3\, +},\, q_{2\, -},\, q_{1\, -}},
\end{eqnarray}  
i.e. an event which is a combination of all possible events with $q_2 = -1$ и $q_3 = +1$. The values of the other $q_i$ may be $+1$ or $-1$. The analog of the probability $w(q_{3\, +},\, q_{2\, -}\, |\, t_3,\, t_2,\, t_1)$ can be writen as:
\begin{eqnarray}
&& w(q_{3\, +},\, q_{2\, -}\, |\, t_n,\, t_{n-1},\,\ldots\,\, t_1) = \\
&& =
\sum\limits_{\omega^{(LGn)}_{32} \in\, {\cal K}^{(LGn)}_{32}}\,\sum\limits_{q_n}\,\ldots\,\sum\limits_{q_4}\,\sum\limits_{q_1}\, 
w(\omega^{(LGn)}_{32},\, q_n,\,\ldots\, q_4,\, q_{3\, +},\, q_{2\, -},\, q_1\, |\, t_n,\, t_{n-1},\,\ldots\, t_1). \nonumber
\end{eqnarray}
Then, based on (\ref{LGW-1part}), it is possible to write a chain of inequalities:
\begin{eqnarray}
w(q_{n\, +},\, q_{1\, -}\, |\, t_n,\, t_{n-1},\,\ldots\,\, t_1) &\le&
w(q_{n\, +},\, q_{2\, -}\, |\, t_n,\, t_{n-1},\,\ldots\,\, t_1)\, +\, w(q_{2\, +},\, q_{1\, -}\, |\, t_n,\, t_{n-1},\,\ldots\,\, t_1); \nonumber\\
w(q_{n\, +},\, q_{2\, -}\, |\, t_n,\, t_{n-1},\,\ldots\,\, t_1) &\le&
w(q_{n\, +},\, q_{3\, -}\, |\, t_n,\, t_{n-1},\,\ldots\,\, t_1)\, +\, w(q_{3\, +},\, q_{2\, -}\, |\, t_n,\, t_{n-1},\,\ldots\,\, t_1); \nonumber\\
&\ldots& \nonumber\\
w(q_{n\, +},\, q_{(n-2)\, -}\, |\, t_n,\, t_{n-1},\,\ldots\,\, t_1) &\le&
w(q_{n\, +},\, q_{(n-1)\, -}\, |\, t_n,\, t_{n-1},\,\ldots\,\, t_1)\, + \nonumber \\
&+&\, w(q_{(n-1)\, +},\, q_{(n-2)\, -}\, |\, t_n,\, t_{n-1},\,\ldots\,\, t_1). \nonumber
\end{eqnarray}
Using this chain we obtain the generalization of the one-particle inequality (\ref{LGW-1part}) for $n$ times:
\begin{eqnarray}
\label{LGWn-1part}
&&     w(q_{n\, +},\, q_{1\, -}\, |\, t_n,\, t_{n-1},\,\ldots\,\, t_1) \le  \nonumber \\
&\le& w(q_{n\, +},\, q_{(n-1)\, -}\, |\, t_n,\, t_{n-1},\,\ldots\,\, t_1)\, + w(q_{(n-1)\, +},\, q_{(n-2)\, -}\, |\, t_n,\, t_{n-1},\,\ldots\,\, t_1)\, +\,\ldots \\
\ldots& +& w(q_{3\, +},\, q_{2\, -}\, |\, t_n,\, t_{n-1},\,\ldots\,\, t_1)\, +\,  w(q_{2\, +},\, q_{1\, -}\, |\, t_n,\, t_{n-1},\,\ldots\,\, t_1).\nonumber
\end{eqnarray} 
This inequality is defined on the set $\Omega^{(LGn)}$.

The inequality (\ref{LGWn-1part}) is quite easily violated in quantum mechanics. Consider a precession of a spin $s=1/2$ in a constant and homogeneous magnetic field, oriented along the axis $y$. Let us set the field intensity such that during the time $\Delta t = t_n - t_1$, the spin rotates by an angle $\pi$ in the $(x,\, z)$ plane. Let us choose the intervals between times $t_i$ and $t_{i+1}$ to be equal and study the spin projections onto axes defined by unitary vectors $\vec a_i$, lying in $(x,\, z)$ plane. In this case the angle between the vectors $\vec a_{i+1}$ and $\vec a_i$ will be $\theta_{i+1,\, i} = \pi/(n-1)$, while the angle between the vectors $\vec a_n$ and $\vec a_1$ will be $\theta_{n,\, 1} = \pi$. Then the inequality (\ref{LGWn-1part}) may be written as:
$$
\sin^2 \left ( \frac{\pi}{2}\right)\, \le\, (n - 1)\,\sin^2 \left ( \frac{\pi}{2 (n - 1)}\right).
$$
For $n \gg 1$ it transforms into a false inequality 
\begin{eqnarray}
1\,\le\,\frac{\pi^2}{4}\,\frac{1}{n - 1}\, \to\, 0.
\end{eqnarray}
The angle $\theta_{n\, 1}$ may be freely chosen, except $0$ and $2\pi$. In the limit $n \to \infty$ the inequality (\ref{LGWn-1part}) will be violated. So for the spin in the magnetic field any positive probability contradicts the concept of macroscopic realism.  

This statement may be generalized as a theorem: the macroscopic realism concept leads to the fact that the probability $w(q_{n\, +},\, q_{1\, -}\, |\, t_n,\, t_{n-1},\,\ldots\,\, t_1)$, calculated in the framework of quantum mechanics in the limit  $n \to \infty$, cannot be positive. This is the one of the main results of the present work.

The theorem is a corollary of the quantum Zeno paradox \cite{zeno1958,zeno1977}, applied to the macroscopic realism. But we will introduce another proof. 

Let the evolution of a closed quantum system is defined by a Hamiltonian $\hat H$. Using the orthogonality condition 
$\bracket{q_{i\, +}}{q_{(i-1)\, -}} = 0$ we find that
\begin{eqnarray}
\label{LGWn-wiip1}
w(q_{i\, +},\, q_{(i - 1)\, -}\, |\, t_n,\, t_{n-1},\,\ldots\,\, t_1) & =&
\left |
\matrel{q_{i\, +}}{e^{-\,\frac{i}{\hbar}\,\hat H\,\frac{t_n - t_1}{n-1}}}{q_{(i - 1)\, -}} 
\right |^2
\approx \nonumber \\
& \approx &
\frac{(t_n - t_1)^2}{(n-1)^2}\frac{1}{\hbar^2}\,\sigma^{(H)}_{i\, i-1},
\end{eqnarray}
where
$$
\sigma^{(H)}_{i\, i-1} =\matrel{q_{i\, +}}{\hat H}{q_{(i - 1)\, -}}\matrel{q_{(i - 1)\, -}}{\hat H}{q_{i \, +}}\, -\,
\frac{1}{2}\left ( 
\matrel{q_{i\, +}}{\hat H^2}{q_{(i - 1)\, -}} + \matrel{q_{(i-1)\, -}}{\hat H^2}{q_{i \, +}}
\right ).
$$  
From (\ref{LGWn-wiip1}) it follows that all $\sigma^{(H)}_{i\, i-1} \ge 0$. Let us denote the maximal value from the set 
$
\left\{\frac{1}{\hbar^2}\,\sigma^{(H)}_{i\, i-1}\right\}
$
as $\left | {\cal M}\right |^2$. Then from (\ref{LGWn-1part}) and (\ref{LGWn-wiip1}) we obtain that 
$$
w(q_{n\, +},\, q_{1\, -}\, |\, t_n,\, t_{n-1},\,\ldots\,\, t_1) \le  \frac{(t_n - t_1)^2}{n-1}\, \left | {\cal M}\right |^2\, \to\, 0
$$
when $n \to \infty$, because $t_n - t_1$ and $\left | {\cal M}\right |^2$ are finite. Q.E.D.

So one can state that the application of macroscopic realism to quantum sysmems must lead to Zeno effect \cite{zeno1958,zeno1977}, i.e. th freezing of the quantum system in the initial state. This is the simplest experimental test of the concept if macroscopic realism for micro-world. It is known experimentally that closed quantum systems do evolve in time, so one can conclude that the concept of macroscopic realism is not suitable for the description of quantum phenomena.


\section{A logical argument against the concept of local realism}
\label{sec:LGI-02a}

Note that in analogy it is possible to introduce a statement about the Wigner inequality for a single particle. For instance (all the notations are defined in Appendix \ref{sec:Awlg}):
\begin{eqnarray}
\label{Wigner_n}
  \lim\limits_{n \to \infty}  w(a_+,\, b_-\, |\, A,\, C_{n-1},\, C_{n-2},\,\,\ldots\,\, C_2,\, B)\,\le\, 0. 
\end{eqnarray}

It means that if a particle has an infinite set of distinct observables then the joint probability of existing of any two of its observables is zero. It seems this statement also contradicts to quantum mechanics, as if the operators of the observables $A$ and $B$ commute, then this joint probability may not be zero. I.e. we present yet another proof of the impossibility of combination of the CR (with NSC) and quantum physics. This fact may be considered as an additional argument of the incompatibility of the LR concept with the principles of quantum mechanics.


\section{Test of the hypothesis of realism}
\label{sec:LGI-03}

Let us consider a closed physical system which consists of two subsystems, ``$1$'' and ``$2$''. In each of the subsystems there is a dichotomic variable $Q^{(\eta)}(t)$, where $\eta =\{1,\, 2\}$ is the subsystem index. Let us consider three times, $t_3 > t_2 > t_1$. At time $t_1$ there is an anticorrelation between dichotomic variables $Q^{(1)}(t)$ and $Q^{(2)}(t)$ like $Q^{(1)}(t_1) =  - Q^{(2)}(t_1)$, or
\begin{eqnarray}
\label{q-anticorrelation}
q^{(1)}_{1\,\pm} =\, -\, q^{(2)}_{1\,\mp}.
\end{eqnarray}
If at time $t_1$ a measurement of $Q^{(\eta)}(t_1)$ occured, then at times $t_2$ and $t_3$ there is no correlation between $Q^{(1)}(t)$ and $Q^{(2)}(t)$. If at time $t_1$ there is no measurement of $Q^{(\eta)}(t_1)$, then the anticorrelation (\ref{q-anticorrelation}) will hold at $t_2$. Note, that by definition at $t_3$ the anticorrelation between the observables $Q^{(1)}(t)$ and $Q^{(2)}(t)$ cannot be observed under any conditions.

We introduce a space of elementary outcomes $\omega^{(\widetilde{LG})} \in \Omega^{(\widetilde{LG})}$, which consists of the aggregates 
$$
\{ q^{(2)}_{3\,\alpha},\, q^{(2)}_{2\,\beta},\, q^{(2)}_{1\,\gamma},\, q^{(1)}_{3\,\alpha'},\, q^{(1)}_{2\,\beta'},\, q^{(1)}_{1\,\gamma' = -\gamma}\},
$$
where the indices $\{ \alpha,\, \beta,\, \gamma,\, \alpha',\, \beta',\,\gamma'\} = \{ +,\, -\}$, and the anticorrelation condition (\ref{q-anticorrelation}) is taken into account. Denote an elementary event as:
$
\protect{\cal K}^{(\widetilde{LG})}_{q^{(2)}_{3\,\alpha},\, q^{(2)}_{2\,\beta},\, q^{(2)}_{1\,\gamma},\, q^{(1)}_{3\,\alpha'},\, q^{(1)}_{2\,\beta'},\, q^{(1)}_{1\,\gamma' = -\gamma}} \subseteq \Omega^{(\widetilde{LG})}
$. The full aggregate of such events forms a $\sigma$--algebra ${\cal F}^{(\widetilde{LG})}$. On
$
\left (\Omega^{(\widetilde{LG})},\, {\cal F}^{(\widetilde{LG})}\right )  
$
let us introduce a non-negative $\sigma$--additive measure 
$
w \left ( \omega^{(\widetilde{LG})},\, q^{(2)}_{3\,\alpha},\, q^{(2)}_{2\,\beta},\, q^{(2)}_{1\,\gamma},\, q^{(1)}_{3\,\alpha'},\, q^{(1)}_{2\,\beta'},\, q^{(1)}_{1\, \gamma'  = -\gamma}\, |\, t_3,\, t_2,\, t_1 \right )
$.
The triplet 
$
\left (\Omega^{(\widetilde{LG})},\, {\cal F}^{(\widetilde{LG})},\, w(\ldots) \right )  
$
is a probabilistic model, which will be used to test the hypothesis of realism.

For rigorous application in the framework of Kolmogorov axiomatics, the mathematical form of the no-signaling in time condition (\ref{NSIT}) should be corrected for the definition of the elementary outcome as follows:
\begin{eqnarray}
\label{NISTomega}
\sum\limits_{\omega^{(\widetilde{LG})}_{ij \ldots} \in\, {\cal K}^{(\widetilde{LG})}_{ij \ldots}}
\sum\limits_{q_k}\, w(\omega^{(\widetilde{LG})}_{ij \ldots},\, q_j,\, q_k,\, q_i,\, \ldots\, |\, t_j,\, t_k,\, t_i,\, \ldots )\, =\, w(q_j,\, q_i,\, \ldots\, |t_j,\, t_i,\, \ldots).
\end{eqnarray}   

We now prove the inequality which is analogous to Wigner inequality (\ref{wigner_nastoyaschiy_wigner}). We introduce an event:
\begin{eqnarray}
\label{K3+2+LG}
{\cal K}^{(\widetilde{LG})}_{32} &=&
{\cal K}^{(\widetilde{LG})}_{q^{(2)}_{3+},\, q^{(2)}_{2-},\, q^{(2)}_{1+},\, q^{(1)}_{3+},\, q^{(1)}_{2+},\, q^{(1)}_{1-}}\,\cup\,
{\cal K}^{(\widetilde{LG})}_{q^{(2)}_{3+},\, q^{(2)}_{2-},\, q^{(2)}_{1+},\, q^{(1)}_{3-},\, q^{(1)}_{2+},\, q^{(1)}_{1-}}\,\cup \\
&\cup&
{\cal K}^{(\widetilde{LG})}_{q^{(2)}_{3+},\, q^{(2)}_{2-},\, q^{(2)}_{1-},\, q^{(1)}_{3+},\, q^{(1)}_{2+},\, q^{(1)}_{1+}}\,\cup\,
{\cal K}^{(\widetilde{LG})}_{q^{(2)}_{3+},\, q^{(2)}_{2-},\, q^{(2)}_{1-},\, q^{(1)}_{3-},\, q^{(1)}_{2+},\, q^{(1)}_{1+}}\, .\nonumber
\end{eqnarray}
Equation (\ref{K3+2+LG}) takes into account that the variables $Q^{(1)}(t)$ and $Q^{(2)}(t)$ are anticorrelated at time $t_1$, as well as at time $t_2$, because there has been no measurement at $t_1$. Then, taking into account (\ref{NISTomega}), we may write: 
\begin{eqnarray}
\label{W3+2+LG}
w \left ( q^{(2)}_{3+},\, q^{(1)}_{2+}\, |\, t_3,\, t_2 \right )  &=&
\sum\limits_{\omega^{(\widetilde{LG})}_{32} \in\, {\cal K}^{(\widetilde{LG})}_{32}}
\sum\limits_{q_3^{(1)}}\, \sum\limits_{q_1^{(1)}}\,\sum\limits_{q_1^{(2)}}\, \delta_{- q_1^{(1)}\,\, q_1^{(2)}} \\
&&
w \left ( \omega^{(\widetilde{LG})}_{32},\, q^{(2)}_{3\, +},\, q^{(2)}_{2\, -},\, q^{(2)}_{1},\, q^{(1)}_{3},\, q^{(1)}_{2\, +},\, q^{(1)}_{1}\, |\, t_3,\, t_2,\, t_1 \right ), \nonumber
\end{eqnarray}
where $\delta_{ij}$ is the Kronecker delta. Then let us introduce another two events. These are: 
\begin{eqnarray}
\label{K3+1+LG}
{\cal K}^{(\widetilde{LG})}_{31} &=&
{\cal K}^{(\widetilde{LG})}_{q^{(2)}_{3+},\, q^{(2)}_{2+},\, q^{(2)}_{1-},\, q^{(1)}_{3+},\, q^{(1)}_{2+},\, q^{(1)}_{1+}}\,\cup\,
{\cal K}^{(\widetilde{LG})}_{q^{(2)}_{3+},\, q^{(2)}_{2+},\, q^{(2)}_{1-},\, q^{(1)}_{3+},\, q^{(1)}_{2-},\, q^{(1)}_{1+}}\,\cup \nonumber \\
&\cup&
{\cal K}^{(\widetilde{LG})}_{q^{(2)}_{3+},\, q^{(2)}_{2+},\, q^{(2)}_{1-},\, q^{(1)}_{3-},\, q^{(1)}_{2+},\, q^{(1)}_{1+}}\,\cup\,
{\cal K}^{(\widetilde{LG})}_{q^{(2)}_{3+},\, q^{(2)}_{2+},\, q^{(2)}_{1-},\, q^{(1)}_{3-},\, q^{(1)}_{2-},\, q^{(1)}_{1+}} \,\cup \\
&\cup&
{\cal K}^{(\widetilde{LG})}_{q^{(2)}_{3+},\, q^{(2)}_{2-},\, q^{(2)}_{1-},\, q^{(1)}_{3+},\, q^{(1)}_{2+},\, q^{(1)}_{1+}}\,\cup\,
{\cal K}^{(\widetilde{LG})}_{q^{(2)}_{3+},\, q^{(2)}_{2-},\, q^{(2)}_{1-},\, q^{(1)}_{3+},\, q^{(1)}_{2-},\, q^{(1)}_{1+}}\,\cup \nonumber \\
&\cup&
{\cal K}^{(\widetilde{LG})}_{q^{(2)}_{3+},\, q^{(2)}_{2-},\, q^{(2)}_{1-},\, q^{(1)}_{3-},\, q^{(1)}_{2+},\, q^{(1)}_{1+}}\,\cup\,
{\cal K}^{(\widetilde{LG})}_{q^{(2)}_{3+},\, q^{(2)}_{2-},\, q^{(2)}_{1-},\, q^{(1)}_{3-},\, q^{(1)}_{2-},\, q^{(1)}_{1+}}  \nonumber
\end{eqnarray}
and
\begin{eqnarray}
\label{K1+2+LG}
{\cal K}^{(\widetilde{LG})}_{12} &=&
{\cal K}^{(\widetilde{LG})}_{q^{(2)}_{3+},\, q^{(2)}_{2+},\, q^{(2)}_{1+},\, q^{(1)}_{3+},\, q^{(1)}_{2+},\, q^{(1)}_{1-}}\,\cup\,
{\cal K}^{(\widetilde{LG})}_{q^{(2)}_{3+},\, q^{(2)}_{2+},\, q^{(2)}_{1+},\, q^{(1)}_{3-},\, q^{(1)}_{2+},\, q^{(1)}_{1-}}\,\cup \nonumber \\
&\cup&
{\cal K}^{(\widetilde{LG})}_{q^{(2)}_{3+},\, q^{(2)}_{2-},\, q^{(2)}_{1+},\, q^{(1)}_{3+},\, q^{(1)}_{2+},\, q^{(1)}_{1-}}\,\cup\,
{\cal K}^{(\widetilde{LG})}_{q^{(2)}_{3+},\, q^{(2)}_{2-},\, q^{(2)}_{1+},\, q^{(1)}_{3-},\, q^{(1)}_{2+},\, q^{(1)}_{1-}} \,\cup \\
&\cup&
{\cal K}^{(\widetilde{LG})}_{q^{(2)}_{3-},\, q^{(2)}_{2+},\, q^{(2)}_{1+},\, q^{(1)}_{3+},\, q^{(1)}_{2+},\, q^{(1)}_{1-}}\,\cup\,
{\cal K}^{(\widetilde{LG})}_{q^{(2)}_{3-},\, q^{(2)}_{2+},\, q^{(2)}_{1+},\, q^{(1)}_{3-},\, q^{(1)}_{2+},\, q^{(1)}_{1-}}\,\cup \nonumber \\
&\cup&
{\cal K}^{(\widetilde{LG})}_{q^{(2)}_{3-},\, q^{(2)}_{2-},\, q^{(2)}_{1+},\, q^{(1)}_{3+},\, q^{(1)}_{2+},\, q^{(1)}_{1-}}\,\cup\,
{\cal K}^{(\widetilde{LG})}_{q^{(2)}_{3-},\, q^{(2)}_{2-},\, q^{(2)}_{1+},\, q^{(1)}_{3-},\, q^{(1)}_{2+},\, q^{(1)}_{1-}}. \nonumber
\end{eqnarray}
Equations (\ref{K3+1+LG}) and (\ref{K1+2+LG}) take into account that the anticorrelation between $Q^{(1)}(t)$ and $Q^{(2)}(t)$ exists only at the time $t_1$, when the first measurement of one of the observables takes place. This distinguishes (\ref{K3+1+LG}) and (\ref{K1+2+LG}) from (\ref{K3+2+LG}).

For events (\ref{K3+1+LG}) and (\ref{K1+2+LG}) we define probabilities 
\begin{eqnarray}
\label{W3+1+LG}
w \left ( q^{(2)}_{3+},\, q^{(1)}_{1+}\, |\, t_3,\, t_1 \right )  &=&
\sum\limits_{\omega^{(\widetilde{LG})}_{31} \in\, {\cal K}^{(\widetilde{LG})}_{31}}
\sum\limits_{q_3^{(1)}}\, \sum\limits_{q_2^{(1)}}\,\sum\limits_{q_2^{(2)}}\, \\
&&
w \left ( \omega^{(\widetilde{LG})}_{31},\, q^{(2)}_{3\, +},\, q^{(2)}_{2},\, q^{(2)}_{1\, -},\, q^{(1)}_{3},\, q^{(1)}_{2},\, q^{(1)}_{1\, +}\, |\, t_3,\, t_2,\, t_1 \right ), \nonumber
\end{eqnarray}
and
\begin{eqnarray}
\label{W1+2+LG}
w \left ( q^{(2)}_{1+},\, q^{(1)}_{2+}\, |\, t_2,\, t_1 \right )  &=&
\sum\limits_{\omega^{(\widetilde{LG})}_{12} \in\, {\cal K}^{(\widetilde{LG})}_{12}}
\sum\limits_{q_2^{(2)}}\, \sum\limits_{q_3^{(2)}}\,\sum\limits_{q_3^{(1)}}\, \\
&&
w \left ( \omega^{(\widetilde{LG})}_{12},\, q^{(2)}_{3},\, q^{(2)}_{2},\, q^{(2)}_{1\, +},\, q^{(1)}_{3},\, q^{(1)}_{2\, +},\, q^{(1)}_{1\, -}\, |\, t_3,\, t_2,\, t_1 \right ). \nonumber
\end{eqnarray}
The sums (\ref{W3+1+LG}) and (\ref{W1+2+LG}) are defined for the event 
$
{\cal K}^{(\widetilde{LG})}_{321} = {\cal K}^{(\widetilde{LG})}_{31} \cup {\cal K}^{(\widetilde{LG})}_{12}
$. 
This event also contains the event ${\cal K}^{(\widetilde{LG})}_{32}$. Taking into account the non-negativity of the probability measure from (\ref{W3+2+LG}), (\ref{W3+1+LG}), and (\ref{W1+2+LG}) we find that for event ${\cal K}^{(\widetilde{LG})}_{321}$ the following is satisfied: 
\begin{eqnarray}
\label{LGW-2part}
w \left ( q^{(2)}_{3+},\, q^{(1)}_{2+}\, |\, t_3,\, t_2 \right )\,\le\,
w \left ( q^{(2)}_{3+},\, q^{(1)}_{1+}\, |\, t_3,\, t_1 \right )\, +\,
w \left ( q^{(2)}_{1+},\, q^{(1)}_{2+}\, |\, t_2,\, t_1 \right ).
\end{eqnarray}
Inequality (\ref{LGW-2part}) is the main result of the current work. It is obtained using the hypothesis of realism and the no-signaling in time condition. In quantum mechanics the inequality (\ref{LGW-2part}) is violated in the same way as the inequality (\ref{LGW-1part}).

Consider an example of violation of (\ref{LGW-2part}) in quantum mechanics. We will use notations and calculation technique from Appendix ~\ref{sec:B-time-evolution}. Consider a pair of neutral pseudoscalar mesons, which at time $t_1 = 0$ are in Bell-entangled state (\ref{Bell-PsiPlus}).  This state is anticorrelated by flavor of the pair, but is correlated by $CP$-parity and mass/lifetime. Hence this state can not be used for test of Wigner inequality~(\ref{wigner_nastoyaschiy_wigner}), but can be used for test of the inequality ~(\ref{LGW-2part}).

Let us choose as an onservable $Q^{(\eta)}(t)$ the flavor of pseudoscalar meson. $Q = +1$, corresponds to meson with flavor ``M'', while $Q=-1$ -- to meson flavor ``$\bar M$''. Using the relation for the state vector (\ref{Psi-1}), the probability $w \left ( q^{(2)}_{3+},\, q^{(1)}_{2+}\, |\, t_3,\, t_2 \right )$ may be written as:
\begin{eqnarray}
\label{WW-1}
w \left ( q^{(2)}_{3+},\, q^{(1)}_{2+}\, |\, t_3,\, t_2 \right ) &=&
\frac{1}{4}\, e^{- 2 \Gamma t_3}\,\mathrm{ch}\,\left (\frac{\Delta \Gamma\,\Delta t_{32}}{2}\right ) \\
&&
\left [
\mathrm{ch}\,\left (\frac{\Delta \Gamma\, (t_2 + t_3)}{2}\right )\, -\, \cos \left ( \Delta m\, (t_2 + t_3)\right )
\right ].
\nonumber
\end{eqnarray}
In analogy, using (\ref{Psi-2}) and (\ref{Psi-3}) we obtain:
\begin{eqnarray}
\label{WW-2}
w \left ( q^{(2)}_{3+},\, q^{(1)}_{1+}\, |\, t_3,\, t_1 \right ) &=&
\frac{1}{4}\, e^{- 2 \Gamma t_3}\,\mathrm{ch}\,\left (\frac{\Delta \Gamma\,\Delta t_3}{2}\right )\,
\left [
\mathrm{ch}\,\left (\frac{\Delta \Gamma\,  t_3}{2}\right )\, -\, \cos \left ( \Delta m\, t_3 \right )
\right ]
\end{eqnarray}
and
\begin{eqnarray}
\label{WW-3}
w \left ( q^{(2)}_{1+},\, q^{(1)}_{2+}\, |\, t_2,\, t_1 \right )&=& 
\frac{1}{4}\, e^{- 2 \Gamma t_3}\,\mathrm{ch}\,\left (\frac{\Delta \Gamma\,\Delta t_{32}}{2}\right )\,
                                                        \mathrm{ch}\,\left (\frac{\Delta \Gamma\, t_3}{2}\right ) \\
&&
\left [
\mathrm{ch}\,\left (\frac{\Delta \Gamma\,  t_2}{2}\right )\, -\, \cos \left ( \Delta m\, t_2 \right )
\right ].
\nonumber
\end{eqnarray}
Denote
$$
\kappa\, =\,\frac{\Delta\,\Gamma}{2\, \Delta\, m},\qquad \alpha\, =\,\Delta m\, t_3,\qquad \beta\, =\, \Delta m\, t_2.
$$
Then substituting (\ref{WW-1}) -- (\ref{WW-3}) into (\ref{LGW-2part}) leads to the following inequality:
\begin{eqnarray}
\label{LGW-2part-WW}
&&\Big [ \mathrm{ch} (\kappa\, (\alpha + \beta))\, -\, \cos (\alpha + \beta)\Big ]\,\mathrm{ch} (\kappa\, (\alpha - \beta))\,\le \nonumber \\
&\le& 
\Big [ \mathrm{ch} (\kappa\,\alpha)\, -\, \cos (\alpha)\Big ]\,\mathrm{ch} (\kappa\,\alpha)\, +\,
\Big [ \mathrm{ch} (\kappa\,\beta)\, -\, \cos (\beta)\Big ]\,\mathrm{ch} (\kappa\, (\alpha - \beta))\,\mathrm{ch} (\kappa\,\alpha)
\end{eqnarray}
In order to simplify the above inequality let us consider $B_s\,\bar B_s$--meson pairs. For $B_s$--meson $\Delta\Gamma \approx -6.0 \times 10^{-11}$ MeV and $\Delta m \approx 1.2 \times 10^{-8}$ MeV \cite{Nikitin:2015bca}. Hence $\kappa \approx - 2.5 \times 10^{-3}$. I.e. violation of (\ref{LGW-2part-WW}) may be considered in $\kappa = 0$ regime. In this case inequality (\ref{LGW-2part-WW}) turns into simple relation:
\begin{eqnarray}
\label{LGW-2part-simplest}
 \cos (\alpha)\, +\, \cos (\beta)\, -\, \cos (\alpha + \beta)\,\le\, 1
\end{eqnarray}
for $\alpha > \beta > 0$. Choose $\displaystyle\alpha = \frac{3 \pi}{8}$ and $\displaystyle\beta = \frac{3 \pi}{10}$.  Then $\cos \alpha \approx = 0.383$, $\cos\beta \approx 0.588$, and $\cos (\alpha + \beta) \approx -0.522$,  which leads to violation of inequality (\ref{LGW-2part-simplest}), and consequently to violation of inequality (\ref{LGW-2part}). We have shown that inequality(\ref{LGW-2part}) may be violated in quantum theory.

It might appear to be possible to introduce a one-to-one correspondence between inequality (\ref{LGW-2part}) and Wigner inequality (\ref{wigner_nastoyaschiy_wigner}) such as $a^{(\eta)}_{\pm} \to q^{(\eta)}_{3\, \pm}$, $b^{(\eta)}_{\pm} \to q^{(\eta)}_{2\, \pm}$, and $c^{(\eta)}_{\pm} \to q^{(\eta)}_{1\, \pm}$, where $\eta = \{1,\, 2\}$. But this is not true. 

Instead, let us apply to (\ref{wigner_nastoyaschiy_wigner}) a cyclic permutation $a^{(\eta)}_{\pm} \to b^{(\eta)}_{\pm}$,  $b^{(\eta)}_{\pm} \to c^{(\eta)}_{\pm}$, and $c^{(\eta)}_{\pm} \to a^{(\eta)}_{\pm}$. Inequality (\ref{wigner_nastoyaschiy_wigner}) will transform to a new inequality 
\begin{eqnarray}
\label{wigner_nastoyaschiy_wigner_ciklicheskiy}
w(b^{(2)}_+,\, c^{(1)}_+\, |\, B^{(2)},\, C^{(1)})\,\le\, w(a^{(2)}_+,\, c^{(1)}_+\, |\, A^{(2)},\, C^{(1)})\, +\, w(b^{(2)}_+,\, a^{(1)}_+\, |\, B^{(2)},\, A^{(1)}).
\end{eqnarray}  
For the proof of (\ref{wigner_nastoyaschiy_wigner_ciklicheskiy}) on the set of elementary outcomes $\tilde\Omega$, which is defined in Appendix~\ref{sec:Awlg}, it is necessary to introduce three events:
\begin{eqnarray}
\tilde{\mathcal{B}} &=& 
\mathcal{K}_{a^{(1)}_+ b^{(1)}_- c^{(1)}_+\, 
             a^{(2)}_- b^{(2)}_+ c^{(2)}_-}\,\cup\,
\mathcal{K}_{a^{(1)}_- b^{(1)}_- c^{(1)}_+\, 
             a^{(2)}_+ b^{(2)}_+ c^{(2)}_-}\,\subseteq\, \tilde\Omega,\nonumber \\
\mathcal{C} &=& 
\mathcal{K}_{a^{(1)}_- b^{(1)}_+ c^{(1)}_+\, 
             a^{(2)}_+ b^{(2)}_- c^{(2)}_-}\,\cup\,
\mathcal{K}_{a^{(1)}_- b^{(1)}_- c^{(1)}_+\, 
             a^{(2)}_+ b^{(2)}_+ c^{(2)}_-}\,\subseteq\, \tilde\Omega,\nonumber \\
\tilde{\mathcal{A}} &=& 
\mathcal{K}_{a^{(1)}_+ b^{(1)}_- c^{(1)}_+\, 
             a^{(2)}_- b^{(2)}_+ c^{(2)}_-}\,\cup\,
\mathcal{K}_{a^{(1)}_+ b^{(1)}_- c^{(1)}_-\, 
             a^{(2)}_- b^{(2)}_+ c^{(2)}_+}\,\subseteq\, \tilde\Omega\nonumber 
\end{eqnarray}
and repeat all the steps used in the proof of (\ref{wigner_nastoyaschiy_wigner}).
The probability $w(b^{(2)}_+,\, c^{(1)}_+\, |\, B^{(2)},\, C^{(1)})$ is defined for the event $\tilde{\mathcal{B}}\subseteq \tilde{\mathcal{A}} \cup \mathcal{C}$, and the sum of probabilities $w(a^{(2)}_+,\, c^{(1)}_+\, |\, A^{(2)},\, C^{(1)})$ и $w(b^{(2)}_+,\, a^{(1)}_+\, |\, B^{(2)},\, A^{(1)})$ is defined for the event $\tilde{\mathcal{A}} \cup \mathcal{C}$. Hence (\ref{wigner_nastoyaschiy_wigner_ciklicheskiy}) is valid for the event $\tilde{\mathcal{A}} \cup \mathcal{C}$.

For inequality (\ref{LGW-2part}), one might examine whether an analogous cyclic permutation may be introduced: $q^{(\eta)}_{3\, \pm} \to q^{(\eta)}_{2\, \pm}$, $q^{(\eta)}_{2\, \pm} \to q^{(\eta)}_{1\, \pm}$, and $q^{(\eta)}_{1\, \pm} \to q^{(\eta)}_{3\, \pm}$. However after this permutation the valid inequality (\ref{LGW-2part}) transforms into a false inequality:
\begin{eqnarray}
\label{NoLGW-2part}
w \left ( q^{(2)}_{2+},\, q^{(1)}_{1+}\, |\, t_2,\, t_1 \right )\,\le\,
w \left ( q^{(2)}_{2+},\, q^{(1)}_{3+}\, |\, t_3,\, t_2 \right )\, +\,
w \left ( q^{(2)}_{3+},\, q^{(1)}_{1+}\, |\, t_3,\, t_1 \right ),
\end{eqnarray}
because in the space of elementary outcomes $\Omega^{(\widetilde{LG})}$ there are no events for which the left and the right part of the inequality are simultaneously true. The left side of inequality (\ref{NoLGW-2part}) is valid for the event 
\begin{eqnarray}
{\cal K}^{(\widetilde{LG})}_{21} &=&
{\cal K}^{(\widetilde{LG})}_{q^{(2)}_{3+},\, q^{(2)}_{2+},\, q^{(2)}_{1-},\, q^{(1)}_{3+},\, q^{(1)}_{2+},\, q^{(1)}_{1+}}\,\cup\,
{\cal K}^{(\widetilde{LG})}_{q^{(2)}_{3+},\, q^{(2)}_{2+},\, q^{(2)}_{1-},\, q^{(1)}_{3-},\, q^{(1)}_{2+},\, q^{(1)}_{1+}}\,\cup \nonumber \\
&\cup&
{\cal K}^{(\widetilde{LG})}_{q^{(2)}_{3+},\, q^{(2)}_{2+},\, q^{(2)}_{1-},\, q^{(1)}_{3+},\, q^{(1)}_{2-},\, q^{(1)}_{1+}}\,\cup\,
{\cal K}^{(\widetilde{LG})}_{q^{(2)}_{3+},\, q^{(2)}_{2+},\, q^{(2)}_{1-},\, q^{(1)}_{3-},\, q^{(1)}_{2-},\, q^{(1)}_{1+}} \,\cup \nonumber \\
&\cup&
{\cal K}^{(\widetilde{LG})}_{q^{(2)}_{3-},\, q^{(2)}_{2+},\, q^{(2)}_{1-},\, q^{(1)}_{3+},\, q^{(1)}_{2+},\, q^{(1)}_{1+}}\,\cup\,
{\cal K}^{(\widetilde{LG})}_{q^{(2)}_{3-},\, q^{(2)}_{2+},\, q^{(2)}_{1-},\, q^{(1)}_{3-},\, q^{(1)}_{2+},\, q^{(1)}_{1+}}\,\cup \nonumber \\
&\cup&
{\cal K}^{(\widetilde{LG})}_{q^{(2)}_{3-},\, q^{(2)}_{2+},\, q^{(2)}_{1-},\, q^{(1)}_{3+},\, q^{(1)}_{2-},\, q^{(1)}_{1+}}\,\cup\,
{\cal K}^{(\widetilde{LG})}_{q^{(2)}_{3-},\, q^{(2)}_{2+},\, q^{(2)}_{1-},\, q^{(1)}_{3-},\, q^{(1)}_{2-},\, q^{(1)}_{1+}}. \nonumber
\end{eqnarray}
And at the same time the right side of inequality (\ref{NoLGW-2part}) is valid for the event 
$
{\cal K}^{(\widetilde{LG})}_{23} \cup {\cal K}^{(\widetilde{LG})}_{31}
$,
where event ${\cal K}^{(\widetilde{LG})}_{31}$ is defined in (\ref{K3+1+LG}), and the event 
\begin{eqnarray}
{\cal K}^{(\widetilde{LG})}_{23} &=&
{\cal K}^{(\widetilde{LG})}_{q^{(2)}_{3+},\, q^{(2)}_{2+},\, q^{(2)}_{1+},\, q^{(1)}_{3+},\, q^{(1)}_{2-},\, q^{(1)}_{1-}}\,\cup\,
{\cal K}^{(\widetilde{LG})}_{q^{(2)}_{3+},\, q^{(2)}_{2+},\, q^{(2)}_{1+},\, q^{(1)}_{3-},\, q^{(1)}_{2-},\, q^{(1)}_{1-}}\,\cup \nonumber \\
&\cup&
{\cal K}^{(\widetilde{LG})}_{q^{(2)}_{3+},\, q^{(2)}_{2+},\, q^{(2)}_{1-},\, q^{(1)}_{3+},\, q^{(1)}_{2-},\, q^{(1)}_{1+}}\,\cup\,
{\cal K}^{(\widetilde{LG})}_{q^{(2)}_{3+},\, q^{(2)}_{2+},\, q^{(2)}_{1-},\, q^{(1)}_{3-},\, q^{(1)}_{2-},\, q^{(1)}_{1+}}\, .\nonumber
\end{eqnarray}
One can see that
$
{\cal K}^{(\widetilde{LG})}_{21}   \not\subseteq {\cal K}^{(\widetilde{LG})}_{23} \cup {\cal K}^{(\widetilde{LG})}_{31}
$. 

There is no equivalence between inequalities (\ref{LGW-2part}) and (\ref{wigner_nastoyaschiy_wigner}), because the space of elementary outcomes $\Omega^{(\widetilde{LG})}$ is not isomorphic to $\tilde\Omega$. This is due to the fact that in $\Omega^{(\widetilde{LG})}$ the anticorrelation between the observables $Q^{(1)}(t)$ and $Q^{(2)}(t)$ takes place at time $t_1$ and sometimes at $t_2$. But in order to establish a one-to-one correspondence between the spaces $\Omega^{(\widetilde{LG})}$ and $\tilde\Omega$ it is necessary for the anticorrelation between $Q^{(1)}(t)$ and $Q^{(2)}(t)$ to hold at any time. In the absence of this equivalence let us hope that the study of hypothesis of realism may provide additional insight into the structure of quantum theory relative to local realism and macroscopic realism concepts.

\section*{Conclusion}

In the present work we obtain an inequality~(\ref{LGW-2part}) for test of the hypothesis of realism. Studying the probabilistic model of this inequality we have shown some fundamental distinctions between the inequality~(\ref{LGW-2part}) and Wigner inequality (\ref{wigner_nastoyaschiy_wigner}). We stress the fact that derivation of~(\ref{LGW-2part}) requires the no-signaling in time condition. We have shown that inequality~(\ref{LGW-2part}) is violated in quantum mechanics.

Basing on Kolmogorov axiomatics of probability theory and the concept of macroscopic realism we present a derivation of the Leggett--Garg inequality in Wigner form for three (\ref{LGW-1part}), (\ref{LGW-1part-NEW}) and for $n$ (\ref{LGWn-1part}) distinct moments of time. We pay special attention to constuction of the probabilistic models  ($\Omega^{(LG)}$, ${\cal F}^{(LG)},\, w(\ldots\, |\,\, t_3,\, t_2,\, t_1 )$) of each of the considered tasks and to the study of the properties of the state-space of each of the inequalities. This distincts present work from the derivation of the corresponding inequalities in \cite{Saha:2015,Kumari:2017}.

Basing on (\ref{LGWn-1part}) we prove a theorem that any unitary evolution in quantum mechanics is not compatible with the macroscopic realism concept, i.e. that the application of the concept of macroscopic realism to the time evolution of micro-particles leads to quantum Zeno paradox.

Inequality~(\ref{Wigner_n}), which is written in analogy with inequality~(\ref{LGWn-1part}) shows that the hypothesis of classical realism for micro-particles may contradict to the possibility of measuring for a single particle of any pair of its observables, even those that are described by commuting operators in quantum mechanics.


\section*{Acknowledgements}

The authors would like to express our deep gratitude to
Prof.~S.~Seidel (University of New Mexico, USA) for help with
preparation of the paper. We would like also to thank
C.~Aleister (Saint~Genis--Pouilly, France) for creating a warm and friendly working
atmosphere for discussions between the authors.

The work was supported by grant 16-12-10280 of the Russian Science
Foundation. One of the authors (N.~Nikitin) expresses his gratitude for
this support.


\newpage
\appendix

\section{The No Signaling Condition and Wigner inequalities}
\label{sec:Awlg}

Using Kolmogorov axiomatics we derive a Wigner inequality for one particle with spin $s = 1/2$. For the no-signaling condition we will obtain a Wigner inequality for a pair of spin anticorrelated fermions. This derivation will shed more light on the role of the no-signaling condition in obtaining the Wigner inequalities. Also this derivation will be needed for comparison of the Wigner inequalities to newly obtained forms of Leggett-Garg inequalities (\ref{LGW-1part}), (\ref{LGW-1part-NEW}) and (\ref{LGW-2part}).

Let us denote as $n_{\pm}$ a state of a particle with a spin projection of $\pm 1/2$ onto an axis defined by a unitary vector $\vec n$. Let us consider spin projections onto three non-parallel axes $\vec a$, $\vec b$, and $\vec c$. Using the classic realism concept it is possible to introduce a space $\Omega$ of elementary outcomes $\omega_i$, consisting of aggregates of spin projections $\{a_\alpha,\, b_\beta,\, c_\gamma \}$, where $\{\alpha,\, \beta,\,\gamma \} = \{+,\, -\}$. We introduce an elementary event ${\cal K}_{a_\alpha,\, b_\beta,\, c_\gamma} \subseteq \Omega$  as a subset of all elementary outcomes $\omega_i \in \Omega$, when the particle simultaneously has spin projections $a_\alpha$, $b_\beta$, and $c_\gamma$ onto $\vec a$, $\vec b$, and $\vec c$ accordingly. The aggregate of events ${\cal K}_{a_\alpha,\, b_\beta,\, c_\gamma}$ forms a $\sigma$--algebra ${\cal F}$. On $\left ( \Omega,\, {\cal F}\right )$ it is possible to introduce a real non-negative $\sigma$--additive measure $w$ for any elementary event. Using this measure it is possible to define joint and conditional probabilities on the set $\Omega$. The triplet $\left ( \Omega,\, {\cal F},\, w\right )$ is a probabilistic model for constructing the Wigner inequalities for a single particle. The aggregate $\{a_\alpha,\, b_\beta,\, c_\gamma \}$ can be thought of as ontic-states \cite{Harrigan2010} of the model. Epistemic-states of the model are the states for which the spin projections are defined for only one or two axes, i.e. states like $\{a_\alpha,\, c_\gamma \}$, or $\{b_\beta\}$.

In (\ref{NSC}) for the no-signaling condition the concept of an elementary outcome $\omega_i$ was not used. This can be fixed if we demand that
 \begin{eqnarray}
\label{NSComega}
\sum\limits_{\omega_i \,\in\, {\cal K}_a\, \subseteq\, \Omega}\,
\sum\limits_a\, w(\omega_i,\, a,\, b_\beta,\, \ldots\, |\, A,\, B,\, \ldots)\, =\, w(b_\beta,\,\ldots\, |\, B,\, \ldots).
\end{eqnarray}
The no-signaling condition in form (\ref{NSComega}) corresponds to Kolmogorov probability theory. Event ${\cal K}_a$ is a combination of all elementary events ${\cal K}_{a_\alpha,\, b_\beta,\, c_\gamma}$, for which $\alpha = \pm$, and index $\beta$ and the rest of the indices have fixed values defined by (\ref{NSComega}).

To obtain the Wigner inequality for a single particle with spin $s=1/2$, let us consider the events 
\begin{eqnarray}
\label{Kabc-def}
{\cal K}_{AB}\, =\, {\cal K}_{a_+,\, b_-,\, c_+}\cup{\cal K}_{a_+,\, b_-,\, c_-}, \quad
{\cal K}_{BC}\, =\, {\cal K}_{a_+,\, b_-,\, c_+}\cup{\cal K}_{a_-,\, b_-,\, c_+}, \quad
{\cal K}_{AC}\, =\, {\cal K}_{a_+,\, b_+,\, c_-}\cup{\cal K}_{a_+,\, b_-,\, c_-} \nonumber
\end{eqnarray}
and elementary outcomes $\omega_i \in {\cal K}_{AB}$, $\omega_j \in {\cal K}_{BC}$, and $\omega_k \in {\cal K}_{AC}$. Given that the spin projections of $\pm 1/2$ onto any of the axes in the framework of classic realism are independent events, we have: 
\begin{eqnarray}
\label{w_ABC}
w(a_+,\, b_-\, |\, A,\, B,\, C) &=&
\sum\limits_{\omega_i \in\, {\cal K}_{AB}}\,\sum\limits_c\, w(\omega_i,\, a_+,\, b_-,\, c\, |\, A,\, B,\, C); \nonumber \\
w(b_-,\, c_+\, |\, A,\, B,\, C) &=&
\sum\limits_{\omega_j \in\, {\cal K}_{BC}}\,\sum\limits_a\, w(\omega_j,\, a,\, b_-,\, c_+\, |\, A,\, B,\, C); \\
w(a_+,\, c_-\, |\, A,\, B,\, C) &=&
\sum\limits_{\omega_k \in\, {\cal K}_{AC}}\,\sum\limits_b\, w(\omega_k,\, a_+,\, b,\, c_-\, |\, A,\, B,\, C). \nonumber
\end{eqnarray}
The sum $w(b_-,\, c_+\, |\, A,\, B,\, C)$ and $w(a_+,\, c_-\, |\, A,\, B,\, C)$ is defined on the set ${\cal K}_{BC} \cup {\cal K}_{AC}\,=\, {\cal K}\,\subseteq\, \Omega$. Given that ${\cal K}_{AB} \subseteq\, {\cal K}$ and $\omega_i \in {\cal K}_{AB}$, then $\omega_i \in {\cal K}$. Taking into account the non-negativity of probabilities in the right side of equations (\ref{w_ABC}), we obtain the Wigner inequality for a single particle: 
\begin{eqnarray}
\label{wigner_takoy_wigner}
w(a_+,\, b_-\, |\, A,\, B,\, C)\,\le\, w(b_-,\, c_+\, |\, A,\, B,\, C)\, +\, w(a_+,\, c_-\, |\, A,\, B,\, C),
\end{eqnarray}  
which is defined on elementary outcomes of the event ${\cal K}$. Note that in the derivation of (\ref{wigner_takoy_wigner}) only the concept of classic realism and Kolmogorov axiomatics are needed. Also, the condition (\ref{w_ABC}) is not identical to no-signaling condition, written in form (\ref{NSComega}), because the devices $A$, $B$, and $C$ measure the same particle, and cannot be separated by space-like intervals. However the inequality (\ref{wigner_takoy_wigner}) is not testable by any setup with a classical measurement device in the final state. 

Practically the Wigner inequalities can be tested if one uses two particles ``$1$'' and ``$2$'' with spins $s=1/2$, if the spin projections onto any axis defined by a unitary vector $\vec n$ fully anticorrelate, i.e. $n^{(1)} = - n^{(2)}$.  Then (\ref{wigner_takoy_wigner}) becomes  
\begin{eqnarray}
\label{wigner_nastoyaschiy_wigner}
w(a^{(2)}_+,\, b^{(1)}_+\, |\, A^{(2)},\, B^{(1)})\,\le\, w(c^{(2)}_+,\, b^{(1)}_+\, |\, C^{(2)},\, B^{(1)})\, +\, w(a^{(2)}_+,\, c^{(1)}_+\, |\, A^{(2)},\, C^{(1)}).
\end{eqnarray}  
The derivation of inequality (\ref{wigner_nastoyaschiy_wigner}) is given in \cite{Nikitin:2014pqa}. This derivation uses some ideas from \cite{khrennikov2000}, which in turn harks back to~\cite{PhysRevLett.48.291} and~\cite{FineJMP23:1982} Here we briefly summarize this derivation is a slightly different notation which is more suitable for comparison of the well-known inequality (\ref{wigner_nastoyaschiy_wigner}) to the obtained new inequality (\ref{LGW-2part}). 

As a first step, define a space $\tilde\Omega$ of elementary outcomes $\omega_i$, which consists of the aggregate of spin projections 
$
\{
a^{(1)}_{\alpha}  b^{(1)}_{\beta}  c^{(1)}_{\gamma}\, 
a^{(2)}_{-\,\alpha} b^{(2)}_{-\,\beta} c^{(2)}_{-\gamma}
\}
$,
which are anticorrelated. The set of such events forms a $\sigma$-algebra $\tilde{\mathcal{F}}$.  On $(\tilde\Omega,\, \tilde{\mathcal{F}})$ we introduce a real non-negative probability measure, which is  $\sigma$--additive. Taking into account the anticorrelation condition, the aggregate of the spin projections in the space $\tilde\Omega$ may be defined by using any triplet of the spin projections onto the axes $\vec a$, $\vec b$, and $\vec c$ despite the particle index ``$1$'' or ``$2$''. That is, the set $\tilde\Omega$ is isomorphic to the space $\Omega$ of the probabilistic model which was constructed during the derivation of inequality (\ref{wigner_takoy_wigner}).

At the second step, define the events 
\begin{eqnarray}
\mathcal{A} &=& 
\mathcal{K}_{a^{(1)}_- b^{(1)}_+ c^{(1)}_+\, 
             a^{(2)}_+ b^{(2)}_- c^{(2)}_-}\,\cup\,
\mathcal{K}_{a^{(1)}_- b^{(1)}_+ c^{(1)}_-\, 
             a^{(2)}_+ b^{(2)}_- c^{(2)}_+}\,\subseteq\, \tilde\Omega,\nonumber \\ 
\mathcal{B} &=& 
\mathcal{K}_{a^{(1)}_- b^{(1)}_+ c^{(1)}_-\, 
             a^{(2)}_+ b^{(2)}_- c^{(2)}_+}\,\cup\,
\mathcal{K}_{a^{(1)}_+ b^{(1)}_+ c^{(1)}_-\, 
             a^{(2)}_- b^{(2)}_- c^{(2)}_+}\,\subseteq\, \tilde\Omega,\nonumber \\
\mathcal{C} &=& 
\mathcal{K}_{a^{(1)}_- b^{(1)}_+ c^{(1)}_+\, 
             a^{(2)}_+ b^{(2)}_- c^{(2)}_-}\,\cup\,
\mathcal{K}_{a^{(1)}_- b^{(1)}_- c^{(1)}_+\, 
             a^{(2)}_+ b^{(2)}_+ c^{(2)}_-}\,\subseteq\, \tilde\Omega.\nonumber 
\end{eqnarray}
The probability $w(a^{(2)}_+,\, b^{(1)}_+\, |\, A^{(2)},\, B^{(1)})$ from formula (\ref{wigner_nastoyaschiy_wigner}) is defined for the event $\mathcal{A}$, and the sum of the probabilities $w(c^{(2)}_+,\, b^{(1)}_+\, |\, C^{(2)},\, B^{(1)})$ and $w(a^{(2)}_+,\, c^{(1)}_+\, |\, A^{(2)},\, C^{(1)})$ are defined for the event $\mathcal{B} \cup \mathcal{C}$, to which the event $\mathcal{A}$ belongs. Hence inequality (\ref{wigner_nastoyaschiy_wigner}) is defined for the event $\mathcal{B} \cup \mathcal{C}$.

At the third step, use instead of the sums (\ref{w_ABC}), the no-signaling condition in form (\ref{NSComega}). This excludes dependence of double probabilities on the third measurement device which can be attributed to subsystem $1$ or to subsystem $2$. Here we will provide more details on the substitution of (\ref{w_ABC}) by the condition (\ref{NSComega}), expanding the explanation in \cite{Nikitin:2014pqa}. The key role in this case is played by the probability 
\begin{eqnarray}
w(a^{(2)}_+,\, c^{(1)}_+\, |\, A^{(2)},\, C^{(1)}) &=&
\sum\limits_{\omega_k \in\, {\cal C}}\,\sum\limits_{b^{(1)}}\, w(\omega_k,\, a^{(2)}_+,\, b^{(1)},\, c^{(1)}_+\, |\, A^{(2)},\, B^{(1)},\, C^{(1)}). \nonumber
\end{eqnarray} 
Given that $\omega_k \in\, {\cal C}$, then under the assumption of locality 
$$
w(\omega_k,\, a^{(2)}_+,\, b^{(1)}_+,\, c^{(1)}_+\, |\, A^{(2)},\, B^{(1)},\, C^{(1)})\, =\,
w(\omega_k,\, a^{(2)}_+,\, b^{(1)}_+,\, c^{(2)}_-\, |\, A^{(2)},\, B^{(1)},\, C^{(2)}),
$$
which, together with the non-negativity of the rest of the probabilities, leads to the inequality (\ref{wigner_nastoyaschiy_wigner}). That very use of the no-signaling condition and locality has allowed us to exclude the dependence of double probabilities on the type of the device $C^{(i)}$ and use the anticorrelation condition. Hence the derivation of the inequality (\ref{wigner_nastoyaschiy_wigner}) is based on the local realism concept and on the no-signaling condition (\ref{NSComega}). Inequality (\ref{wigner_nastoyaschiy_wigner}), unlike inequality (\ref{wigner_takoy_wigner}), can be tested experimentally.

In quantum theory an anticorrelation of spin projections onto any axis corresponds to a spin-singlet maximally entangled Bell state $\ket{\Psi^-}$, which violates the Wigner inequality \cite{wigner}. Given that in the situation of two particles the no-signaling condition is satisfied on both the micro- and the macro-levels, one may state (based on the violation of the Wigner inequalities) that the local realism is violated in quantum mechanics. The question of which of the local realism conditions is violated, locality or the classic realism concept, is still under discussion. In order to exclude locality it is necessary to write the Wigner inequalities in the framework of quantum field theory, which is local by definition \cite{Nikitin:2009sr,Nikitin:2014pqa}.

Instead of the spin $1/2$ projections one can use the quantum numbers of neutral pseudoscalar mesons $B^0$, $B^0_s$, and $K$, which are born in pairs in the entangled state $\ket{\Psi^-}$ in the decays of the $\Upsilon (4S) \to B^0 \bar B^0$, $\Upsilon (5S) \to B^0_s \bar B^0_s$, and $\phi (1020) \to K^0 \bar K^0$.

\section{Description of time evolution of neutral pseudoscalar meson systems}
\label{sec:B-time-evolution}
Consider neutral pseudoscalar mesons $M =\{ K,\, D,\, B_q \}$, where $q = \{ d,\, s\}$. As an observable $Q$ let us choose the flavor of pseudoscalar meson. Let $Q = +1$ for meson flavor $M$ and $Q = -1$ for meson flavor $\bar M$. Let us index each meson of the pair by $\alpha$. I.e. in the pair $\alpha = \{1,\, 2\}$. The state of each of the mesons will be described in two-dimensional Hilbert space ${\cal H}^{(\alpha)}$. Let us introduce in this space a basis of the states with a defined flavor: 
\begin{eqnarray}
\label{flavour-basis}
\ket{M^{(\alpha)}}\, =\,
\left (
\begin{array}{c}
1 \\
0
\end{array}
\right ), \qquad
\ket{{\bar M}^{(\alpha)}}\, =\,
\left (
\begin{array}{c}
0 \\
1
\end{array}
\right ).
\end{eqnarray}  
Also let us define an arbitrary phase of $CP$--conjugation, that 
$$
\hat C\, \hat P\, \ket{M^{(\alpha)}}\, =\, \ket{{\bar M}^{(\alpha)}}, \qquad
\hat C\, \hat P\, \ket{{\bar M}^{(\alpha)}}\, =\,\ket{M^{(\alpha)}}.
$$
In the orthogonal basis (\ref{flavour-basis}) one can write a states with some defined $CP$--parity
\begin{eqnarray}
\ket{M_1^{(\alpha)}} = \frac{1}{\sqrt{2}}\,\left ( \ket{M^{(\alpha)}} +  \ket{\bar M^{(\alpha)}}\right ), \quad 
\ket{M_2^{(\alpha)}} = \frac{1}{\sqrt{2}}\,\left ( \ket{M^{(\alpha)}} - \ket{\bar M^{(\alpha)}}\right ) \nonumber
\end{eqnarray}
and with a defined values of the mass and lifetime
\begin{eqnarray}
\ket{M_L^{(\alpha)}} = p\,\ket{M^{(\alpha)}} + q\, \ket{\bar M^{(\alpha)}}, \quad
\ket{M_H^{(\alpha)}} = p\,\ket{M^{(\alpha)}} - q\, \ket{\bar M^{(\alpha)}}. \nonumber
\end{eqnarray}
$CP$--eigenstates are orthogonal, 
but $\bracket{M_L^{(\alpha)}}{M_H^{(\alpha)}} = |p|^2 - |q|^2 \ne 0$, i.e. in the chosen basis the states with some defined mass and the lifetime are not orthogonal. Complex parameters $p$ and $q$ suit the following normalization condition: 
\begin{eqnarray}
\label{pq-normirovka}
\bracket{M_L^{(\alpha)}}{M_L^{(\alpha)}}=\bracket{M_H^{(\alpha)}}{M_H^{(\alpha)}}=|p|^2 + |q|^2 = 1. 
\end{eqnarray}
Let $\hat D$ to be the operator of converting to basis where the states $\ket{M_{L,\, H}^{(\alpha)}}$ are orthogonal. This operator has the form: 
\begin{eqnarray}
\hat D^{(\alpha)} =\,\frac{1}{2\, p\, q}\,
\left (
\begin{matrix}
q &  p \\
q & -p
\end{matrix}
\right )
\end{eqnarray}
In basis where the states $\ket{M_L^{(\alpha)}}$ and $\ket{M_H^{(\alpha)}}$ are orthogonal, the operator of time evolution has the following form:
 \begin{eqnarray}
\hat U^{(\alpha)}(t)\, =\,
\left (
\begin{matrix}
e^{-i\, E_L t} &  0 \\
0 & e^{-i\, E_H t}
\end{matrix}
\right )
\end{eqnarray}
and
\begin{eqnarray}
E_L  = m_L - \frac{i}{2}\, \Gamma_L;  \qquad E_H = m_H - \frac{i}{2}\, \Gamma_H  
\nonumber
\end{eqnarray}
are the complex energies related to the states $\ket{M_L^{(\alpha)}}$ and $\ket{M_H^{(\alpha)}}$ accordingly. For the subsequent calculations let us use the following definitions: 
\begin{eqnarray}
&&  \Delta m = m_H - m_L, \qquad
\Delta \Gamma = \Gamma_H - \Gamma_L, \qquad
\Gamma\, =\,\frac{1}{2}\,\left ( \Gamma_H + \Gamma_L\right ).
\nonumber
\end{eqnarray}
The inverse operator for $\hat D^{(\alpha)}$, has form: 
\begin{eqnarray}
\left (\hat D^{(\alpha)} \right )^{\, -1} =\,
\left (
\begin{matrix}
p &  p \\
q & -q
\end{matrix}
\right ).
\end{eqnarray}
It changes the orthogonal states $\ket{M_L^{(\alpha)}}$ and $\ket{M_H^{(\alpha)}}$ back into non-orthogonal. Finally in the space ${\cal H}^{(\alpha)}$ let us introduce $S$--matrix 
\begin{eqnarray}
\label{S-matrix}
\hat S^{(\alpha)} (t)\, =\, \left (\hat D^{(\alpha)} \right )^{\, -1} \hat U^{(\alpha)}(t)\,\, \hat D^{(\alpha)}.
\end{eqnarray} 
This matrix satisfy the group property:
\begin{eqnarray}
\label{S12-S1*S2}
\hat S^{(\alpha)} (t_1 + t_2)\, =\,\hat S^{(\alpha)} (t_1)\, \hat S^{(\alpha)} (t_2),
\end{eqnarray}
as the evolution matrix $\hat U^{(\alpha)}(t)$ satisfy it.

In the above technique it is easy to calculate any time evolutions of any states of pseudoscalar mesons. For example:
\begin{eqnarray}
\left \{
\begin{array}{l}
\ket{M^{(\alpha)}(t)}\, =\, \hat S^{(\alpha)} (t)\,  \ket{M^{(\alpha)}}\,
=\, g_+(t)\,\ket{M^{(\alpha)}}\, -\,\frac{q}{p}\, g_-(t)\,\ket{{\bar M}^{(\alpha)}}\\
\ket{{\bar M}^{(\alpha)}(t)}\, =\, \hat S^{(\alpha)} (t)\, \ket{{\bar M}^{(\alpha)}}
=\, g_+(t)\,\ket{{\bar M}^{(\alpha)}}\, -\,\frac{p}{q}\, g_-(t)\,\ket{M^{(\alpha)}}
\end{array}
\right . , \nonumber
\end{eqnarray}
where $\displaystyle g_{\pm}(t) = \frac{1}{2}\,\left ( e^{-i\, E_H t} \pm e^{- i\, E_L t}\right )$.  Functions $g_{\pm}(t)$ satisfy the following conditions:
\begin{eqnarray}
&& \left | g_{\pm}(t) \right |^2 = \frac{e^{-\,\Gamma\, t}}{2}\,
\left (
{\mathrm{ch}} \left ( \frac{\Delta\Gamma\, t}{2}\right ) \pm \cos \left ( \Delta m\, t\right ) \right ), \nonumber \\
&& g_+^* (t)\, g_-(t) \,=\, -\,\frac{e^{-\,\Gamma\, t}}{2}\,
\left (
{\mathrm{sh}} \left ( \frac{\Delta\Gamma\, t}{2}\right ) +  i\sin \left ( \Delta m\, t\right ) 
\right ).  \nonumber 
\end{eqnarray}
Also from the group property (\ref{S12-S1*S2}) it follows that:
\begin{eqnarray}
&& g_+ (t_2 + t_1)\, =\, g_+ (t_2)\, g_+(t_1)\, +\, g_-(t_2)\, g_-(t_1),
\nonumber \\
&& g_-(t_2 + t_1)\, =\, g_+ (t_2)\, g_-(t_1)\, +\, g_-(t_2)\, g_+(t_1).
\nonumber
\end{eqnarray}
Let us consider a pair of pseudoscalar mesons which at time $t=0$ exists in a flavour Bell-entangled state:
\begin{eqnarray}
\label{Bell-PsiPlus}
\ket{\Psi^+}\, =\,\frac{1}{\sqrt{2}}\,
\Big (
\ket{M^{(2)}}\otimes\ket{{\bar M}^{(1)}}\, +\, \ket{{\bar M}^{(2)}}\otimes\ket{M^{(1)}}
\Big )
\end{eqnarray}
Evolution of the state $\ket{\Psi^+}$ is described in Hilbert space ${\cal H} = {\cal H}^{(1)} \otimes {\cal H}^{(2)}$. In this space the $S$--matrix has the form:
$$
\hat S(t)\, =\,\hat S^{(1)} (t)\otimes \hat S^{(2)} (t),
$$
and projectors to states $\ket{M^{(1)}}$ and $\ket{M^{(2)}}$:
$$
{\hat {\cal P}}_M^{(1)} =\, \hat P_M^{(1)} \otimes \hat 1^{(2)}, \qquad
{\hat {\cal P}}_M^{(2)} =\, \hat 1^{(1)} \otimes \hat P_M^{(2)},
$$
where $\hat 1^{(\alpha)}$ -- is a unitary operator in the space ${\cal H}^{(\alpha)}$ and $\hat P_M^{(\alpha)} =\,\ket{M^{(\alpha)}}\bra{M^{(\alpha)}}$ -- is a projector to the state $\ket{M^{(\alpha)}}$ in the space ${\cal H}^{(\alpha)}$.

Consider an example of the above technique. Let in the time $t_1 = 0$ our system was in the state $\ket{\Psi^+}$, then at the time $t_2 > t_1$ the first meson was measured in the state ``$M$'', and at the moment of time $t_3 > t_2$ the second meson was also measured in the state ``$M$''.  Then at time $t_3$ the pair of pseudoscalar mesons is in the state:
\begin{eqnarray}
\label{Psi-1}
\ket{\Psi (t_3,\, t_2,\, [t_1])} &=& {\hat {\cal P}}_M^{(2)}\,\hat S(t_3 - t_2)\, {\hat {\cal P}}_M^{(1)}\,\hat S(t_2 - t_1)\,\ket{\Psi^+ (t_1)}\, =\\
&=& -\,\frac{1}{\sqrt{2}}\,\frac{p}{q}\, g_-(t_2 + t_3)
\left (
g_+(\Delta t_{32})\,\ket{M^{(1)}}\, -\,\frac{q}{p}\, g_-(\Delta t_{32})\,\ket{{\bar M}^{(1)}}
\right ) \otimes \,\ket{M^{(2)}},
\nonumber
\end{eqnarray}
where $\Delta t_{32} = t_3 - t_2$ and $[t_i]$ defines the time where there were no measurements. In order to introduce an example of violation of (\ref{LGW-2part}) from~\ref{sec:LGI-03} let us write two more state vectors. First is related to the fact that at the time $t_1 = 0$ the pair of pseudoscalar mesons was in the state $\ket{\Psi^+}$. Then at $t_1$ the first meson was measured in the state ``$M$'', and then at $t_3 > t_1$ the second meson also was measured in the state ``$M$''.  At the time $t_2$ measurement is not performed. $t_3 > t_2 > t_1$. Then:
\begin{eqnarray}
\label{Psi-2}
\ket{\Psi (t_3,\, [t_2],\, t_1)} &=& {\hat {\cal P}}_M^{(2)}\,\hat S(t_3 - t_2)\,\hat S(t_2 - t_1)\, {\hat {\cal P}}_M^{(1)}\,\ket{\Psi^+ (t_1)}\, =
\nonumber \\
&=&
{\hat {\cal P}}_M^{(2)}\,\hat S(t_3 - t_1)\,{\hat {\cal P}}_M^{(1)}\,\ket{\Psi^+ (t_1)}\, = 
\\
&=& -\,\frac{1}{\sqrt{2}}\,\frac{p}{q}\, g_-(t_3)
\left (
g_+(t_3)\,\ket{M^{(1)}}\, -\,\frac{q}{p}\, g_-(t_3)\,\ket{{\bar M}^{(1)}}
\right ) \otimes \,\ket{M^{(2)}}.
\nonumber
\end{eqnarray}
The second state vector is related to the fact that at $t_1 = 0$ the system was in the state $\ket{\Psi^+}$. At the same time the \textit{second} meson is measured in the state ``$M$''.  At $t_2 > t_1$ the \textit{first} meson is measured in the state ``$M$''. It is necessary to find a state vector at $t_3 > t_2$, where no measurements take place. Have:
\begin{eqnarray}
\label{Psi-3}
\ket{\Psi ([t_3],\,  t_2,\, t_1)} &=& \hat S(t_3 - t_2)\,{\hat {\cal P}}_M^{(1)}\,\hat S(t_2 - t_1)\, {\hat {\cal P}}_M^{(2)}\,\ket{\Psi^+ (t_1)}\, =
\nonumber \\
&=& -\,\frac{1}{\sqrt{2}}\,\frac{p}{q}\, g_-(t_2)
\left (
g_+(\Delta t_{32})\,\ket{M^{(1)}}\, -\,\frac{q}{p}\, g_-(\Delta t_{32})\,\ket{{\bar M}^{(1)}}
\right ) \otimes
\\
& \otimes&
\left (
g_+(t_3)\,\ket{M^{(2)}}\, -\,\frac{q}{p}\, g_-(t_3)\,\ket{{\bar M}^{(2)}}
\right ).
\nonumber
\end{eqnarray}



\newpage
\begin{thebibliography}{99}
\bibitem{Einstein:1935rr}
A.~Einstein, B.~Podolsky, and N.~Rosen, ``Can quantum mechanical description of physical reality be considered complete?'', Phys.\ Rev.\  {\bf 47}, 777  (1935).
\bibitem{Leggett:1985zz}
A.~J.~Leggett and A.~Garg, ``Quantum mechanics versus macroscopic realism: Is the flux there when nobody looks?'', Phys.\ Rev.\ Lett.\  {\bf 54}, 857 (1985).
\bibitem{Leggett:2002}
A.~J.~Leggett,  ``Testing the limits of quantum mechanics: motivation, state of play, prospects'', J.\  Phys.\ Condens.\ Matter  \textbf{14}, R415 (2002).
\bibitem{Leggett:2008} 
A.~J.~Leggett, ``Realism and the physical world'', Rep.\ Prog.\ Phys.\ \textbf{71}, 022001 (2008).
\bibitem{Kofler:2008}
J.~Kofler and C.~Brukner, ``Conditions for quantum violation of macroscopic realism'', Phys.\ Rev.\ Lett.\ \textbf{101}, 090403 (2008).
\bibitem{Cirelson:1980}
B.~S.~Cirel'son, ``Quantum generalizations of Bell's inequality'', Lett.\ Math.\ Phys.\ \textbf{4}, 93 (1980).
\bibitem{PRboxes:1984} 
S.~Popescu and D.~Rohrlich, ``Quantum nonlocality as an axiom'', Found. Phys. \textbf{24}, 379 (1994).
\bibitem{eberhard} 
P.~H.~Eberhard, ``Bell's theorem and the different concepts of locality'', Nuovo Cimento~B\textbf{46}, 392 (1978).
\bibitem{wigner} 
E.~P.~Wigner, ``On hidden variables and quantum mechanical probabilities'', Am.\ J.\ Phys.\ \textbf{38}, 1005 (1970).  
\bibitem{Bell:1964kc} 
 J.~S.~Bell, ``On the Einstein-Podolsky-Rosen paradox'', Physics {\bf 1}, 195 (1964).
\bibitem{Bell:1964fg} 
J.~S.~Bell, ``On the Problem of Hidden Variables in Quantum Mechanics'', Rev.\ Mod.\ Phys.\  {\bf 38}, 447 (1966).
\bibitem{Clauser:1969ny} 
J.~F.~Clauser, M.~A.~Horne, A.~Shimony, and R.~A.~Holt, ``Proposed experiment to test local hidden variable theories'', Phys.\ Rev.\ Lett.\  {\bf 23}, 880 (1969).
\bibitem{PhysRevA.87.052115} 
J.~Kofler and C.~Brukner, ``Condition for macroscopic realism beyond the Leggett-Garg inequalities'', Phys.\  Rev.\ A {\bf 87}, 052115 (2013).
\bibitem{PhysRevA.96.012121}
J. ~J.~ Halliwell,  ``Comparing conditions for macrorealism: Leggett-Garg inequalities versus no-signaling in time'',  Phys.\ Rev.\ A\ \textbf{96}, 012121 (2017).
\bibitem{PhysRevA.54.1798}
S.F.~Huelga, T.W.~Marshall, and E.~Santos, ``Temporal Bell-type inequalities for two-level Rydberg atoms coupled to a high-Q resonator'',
Phys.\ Rev.\ A {\bf 54}, 1798 (1996).
\bibitem{Gangopadhyay:2013aha} 
D.~Gangopadhyay, D.~Home, and A.S.~Roy, ``Probing the Leggett-Garg Inequality for Oscillating Neutral Kaons and Neutrinos'',
Phys.\ Rev.\ A {\bf 88}, 022115 (2013).
\bibitem{Formaggio:2016cuh} 
 J.~A.~Formaggio, D.~I.~Kaiser, M.~M.~Murskyj, and T.~E.~Weiss, ``Violation of the Leggett-Garg Inequality in Neutrino Oscillations'', Phys.\ Rev.\ Lett.\  {\bf 117}, no. 5, 050402 (2016).
\bibitem{Naikoo:2018vug} 
J.~Naikoo, A.K.~Alok, and S.~Banerjee, ``Study of temporal quantum correlations in decohering B and K meson systems'',
Phys.\ Rev.\ D {\bf 97}, 053008 (2018). 
\bibitem{Pusey:2011de}
M.F.~Pusey, J.~Barrett, and T.~Rudolph, ``On the reality of the quantum state'', Nature Phys.\  {\bf 8}, 476 (2012).
\bibitem{Harrigan2010}
N.~Harrigan and R.W.~Spekkens, ``Einstein, Incompleteness, and the Epistemic View of Quantum States'', Found.\ Phys.\  \textbf{40}, 125 (2010).
\bibitem{KAZAKOV20122914}
K.~A.~Kazakov, V.~V.~Nikitin, ``Large-time evolution of an electron in photon bath'', Ann.\ Phys.\ \textbf{327}, pp. 2914-2945 (2012).
\bibitem{Nikitin:2009sr} 
N.~Nikitin and K.~Toms, ``Wigner's inequalities in quantum field theory'', Phys.\ Rev.\ A {\bf 82}, 032109 (2010).
\bibitem{Nikitin:2014pqa} 
N.~Nikitin, V.~Sotnikov, and K.~Toms, ``Time-dependent Bell inequalities in a Wigner form'', Phys.\ Rev.\ A {\bf 90}, 042124 (2014).   
\bibitem{Nikitin:2015bca} 
N.~Nikitin, V.~Sotnikov, and K.~Toms, ``Proposal for experimental test of the time-dependent Wigner inequalities for neutral pseudoscalar meson systems'', Phys.\ Rev.\ D {\bf 92}, no. 1, 016008 (2015).
\bibitem{Nikitin:2016ilm}  
N.~Nikitin and K.~Toms, ``Test of a hypothesis of realism in quantum theory using a bayesian approach'', Phys.\ Rev.\ A {\bf 95}, 052103 (2017).
\bibitem{Saha:2015}
D.~Saha, S.~Mal, P.~K.~Panigrahi, and D.~Home, ``Wigner's form of the Leggett-Garg inequality, the no-signaling-in-time condition, and unsharp measurements'', Phys.\ Rev.\ A  {\bf91}, 032117 (2015).
\bibitem{Mal:2016} 
S.~Mal, D.~Das and D.~Home, ``Quantum mechanical violation of macrorealism for large spin and its robustness against coarse-grained measurements'', Phys.\ Rev.\ A {\bf 94}, 062117 (2016).
\bibitem{Das:2018} 
D.~Das, S.~Mal and D.~Home, ``Testing local-realism and macro-realism under generalized dichotomic measurements'', Phys. Lett.\ A\ {\bf 382}, pp.1085-1091 (2018).
\bibitem{Kumari:2017} 
S.~Kumari and A.~K.~Pan, ``Probing various formulations of macrorealism for unsharp quantum measurements'', Phys.\ Rev.\ A\ {\bf 96}, 042107 (2017).
\bibitem{zeno1958} 
L.~A.~Khalfin, ``Contribution to the decay theory of a quasi-stationary state'', Soviet Phys.\ JETP \textbf{6}, 1053 (1958).
\bibitem{zeno1977}
 E.~C.~G.~Sudarshan, B.~Misra, ``The Zeno's paradox in quantum theory'',  J.\ Math.\ Phys.\ \textbf{18}, 756 (1977).
\bibitem{khrennikov2000}
A.Yu. Khrennikov, ``Non-Kolmogorov probability models and modified Bell's inequality'', J.\, Math.\, Phys., {\bf 41}, 1768 (2000); ``A perturbation of CHSH inequality induced by fluctuations of ensemble distributions'', J.\, Math.\, Phys., {\bf 41}, 5934 (2000).
\bibitem{PhysRevLett.48.291}
A.~Fine, ``Hidden Variables, Joint Probability, and the Bell Inequalities'', Phys.\  Rev.\  Lett.\  \textbf{48}, pp.291--295 (1982).
\bibitem{FineJMP23:1982}
A.~Fine, ``Joint distributions, quantum correlations, and commuting observables'', J.\ Math.\ Phys.\ \textbf{23}, pp.1306--1310 (1982).
\end{thebibliography}
\end{document}